# Quantum Stochastic Approach to the Description of Quantum Measurements

## Elena R. Loubenets[*]

*Moscow State Institute of Electronics and Mathematics,
Trekhsvyatitelskii per. 3/12, Moscow, 109028, Russia*

## Abstract

In the present paper we consider the problem of description of an arbitrary generalized quantum measurement with outcomes in a measurable space.

Analyzing the unitary invariants of a separable statistical realization of a quantum instrument, we present the most general form of an integral representation of an instrument, which differs from the representations of an instrument available in the mathematical and physical literature.

We introduce the notion of a stochastic realization of an instrument and establish a one-to-one correspondence between the class of unitarily and phase equivalent separable statistical realizations and the equivalence class of stochastic realizations of an instrument. We further single out the invariant class of unitarily and phase equivalent separable statistical realizations for which the integral representation of an instrument is the same for all statistical realizations from this class and is wholly determined by the invariants of this class. We call the special form of this integral representation the quantum stochastic representation of an instrument.

We show that the description of a generalized direct quantum measurement can be considered in the frame of a new general approach based on the notion of a family of quantum stochastic evolution operators satisfying the orthonormality relation. This approach gives not only the complete statistical description of any generalized direct quantum measurement but the complete description in a Hilbert space of the stochastic behaviour of a quantum system under a generalized direct measurement in the sense of specifying the probabilistic transition law governing the change from the initial state to a final one under a single measurement. Under this approach a unitary evolution of an isolated quantum system is included as a special case.

In the frame of the proposed approach, which we call quantum stochastic approach, all possible schemes of measurements upon a quantum system can be considered.

In the case of repeated or continuous in time measurements the quantum stochastic approach allows to define, in the most general case, the notion of the family of posterior pure state trajectories (quantum trajectories in discrete or continuous time) in the Hilbert space of a quantum system and to give their probabilistic treatment.

---

# 1.Introduction

The behaviour of an isolated quantum system, which is not observed, is quantum deterministic since it is described by the Schrödinger equation, whose solutions are reversible in time. Under a measurement the behaviour of a quantum system becomes irreversible in time and stochastic. Not only is the outcome of a measured quantum quantity random, being defined with some probability distribution, but the state of the quantum system under a measurement becomes random as well.

We would like to specify from the very beginning that under a quantum measurement we mean a physical experiment upon a quantum system, which resulting in the observation in the classical world of an outcome may cause a change in the state of the quantum system, but not the quantum system's destruction. We distinguish direct and indirect quantum measurements. A direct quantum measurement corresponds to a measurement situation where we have to describe the direct interaction between the measuring device and the observed quantum system, while in case of an indirect measurement, a direct measurement is performed upon some other quantum system, entangled with the one considered.

The term "generalized measurement", as usual, corresponds to the measurement situation with outcomes of the most general nature possible under a quantum measurement.

In quantum measurement theory the formalization of the *complete statistical description* of any generalized quantum measurement is given by the operational approach [2-9]. The complete statistical description implies the knowledge of the probability distribution of different outcomes of the measurement and a statistical description of the state change of the quantum system under the measurement.

However, the operational approach does not, in general, give the possibility to include into consideration the description under a single measurement of the stochastic behaviour of a quantum system, depending on outcomes in the classical world. The description of such stochastic behaviour of a quantum system means the specification of a probabilistic transition law governing the change from the initial state of a quantum system to a final one under a single measurement. We refer to this kind of description as a complete stochastic description of the random behaviour of the quantum system under a single measurement.

The complete stochastic description is, in particular, very important in the case of continuous in time measurements of an open system, where the evolution of the continuously observed open system differs from that described by reversible in time solutions of the Schrödinger equation.

The operational approach also does not, in general, specify the description of a generalized direct quantum measurement.

We would like to underline here that, in general, the description of a direct quantum measurement can not be simply reduced to the quantum theory description of a measuring process. We can not specify definitely neither the interaction, nor the quantum state of a measuring device environment, nor describe a measuring device only in quantum theory terms. In fact, under such a scheme the description of a direct quantum measurement is simply referred to the description of a direct measurement of some observable of an environment of a measuring device. But the problem still remains. Moreover, in quantum theory any physically based problem must be formulated in unitarily equivalent terms and the results of its consideration must not be dependent neither on the choice of a special representation picture (Schrödinger, Heisenberg or interaction) nor on the choice of a basis in the Hilbert space.

We recall that for the case of discrete outcomes the original von Neumann approach [1] in quantum measurement theory describes specifically a direct quantum measurement and gives both – the complete statistical description of a measurement and the complete stochastic description of the random behaviour of the quantum system under a single measurement.

In this paper we present the new mathematical results on the notion of an instrument, which is used in the quantum measurement theory and the theory of open systems. Using these mathematical results, we further introduce a new general approach, the quantum stochastic approach (QSA), to the description of an arbitrary generalized direct quantum measurement based on the physically important mathematical notion of the family of quantum stochastic evolution operators, satisfying the orthonormality relation.

The quantum stochastic approach may be considered as the quantum stochastic generalization of the original von Neumann approach to the description of direct measurements with discrete outcomes to the case of any measurable space of outcomes, any type of a scalar measure on a space of outcomes and any type of a quantum state reduction.

Due to the orthonormality relation, the QSA allows to interpret the posterior pure states, defined by quantum stochastic evolution operators, as posterior pure state outcomes in a Hilbert space corresponding to different random measurement channels. Physically, the notion of different random measurement channels, under the same observed outcome, corresponds, to different underlying random quantum transitions of the environment of a measuring device, which we can not, however, specify with certainty.

In the case when a quantum system is isolated the family of quantum stochastic evolution operators consists of only one element, which is a unitary operator.

The QSA gives not only the complete statistical description of any generalized direct quantum measurement, but it gives alo the complete stochastic description of the random behaviour of the quantum system under a single measurement.

Even for the special case of discrete outcomes, the QSA differs, due to the orthogonality relation for posterior pure state outcomes, from looking somewhat similar approaches considered in the physical literature [18,19], where the so called "measurement" or Kraus operators are used for the description of both the statistics of a measurement (a POV measure) and the conditional state change of a quantum system.

We generalize as far as possible our results presented in [15-17], where the notion of a quantum stochastic operator was defined for the description of conditional evolution of continuously observed quantum systems in the general case of non-demolition measurements.

In Section 2 we review the main approaches to the description of quantum measurements, specifying the characteristic features of each approach.

In Section 3 we present the new mathematical results on the notion of an instrument.

In Section 3.1 we introduce the notion of a class of unitarily and phase equivalent separable statistical realizations of an instrument and find its invariants. In Section 3.2 we present the most general form of an integral representation of an instrument, which differs from the integral representations of an instrument available in the mathematical and physical literature. In Section 3.3 we introduce the notion of a stochastic realization of an instrument and establish a one-to-one correspondence between the class of unitarily and phase equivalent statistical realizations and the equivalence class of stochastic realizations of an instrument. In Section 3.4 we single out invariant classes of unitarily and phase equivalent separable statistical realizations. For the invariant class the integral representation of the corresponding instrument is the same for all statistical realizations from this class and is wholly determined by unitary invariants of a separable statistical realization from this class. We call the special form of the integral representation of an instrument, corresponding to an invariant class, quantum stochastic due to its importance in the quantum measurement theory.

In Section 4 we show that any generalized direct quantum measurement can be interpreted to correspond to an invariant class of unitarily and phase equivalent statistical realizations (measuring processes) and introduce the main ideas of the quantum stochastic approach (QSA). We consider also the description of an indirect quantum measurement in the frame of the QSA.

In Section 5 we give the semiclassical interpretation of the QSA to the description of a generalized direct quantum measurement in terms of the classical probability description of a measuring apparatus and the quantum description of the observed quantum system.

In Section 6 we present the concluding remarks.

## 2. The main approaches to the description of quantum measurements

Let us first review the main approaches to the description of quantum measurements available up to the present moment and specify the characteristic features of each approach.

### 2.1. Von Neumann approach

Let $H_S$ be a complex separable Hilbert space of a quantum system. According to the von Neumann approach [1] only self-adjoint operators on $H_S$ are allowed to represent *real-valued variables* of a quantum system, which can be measured. The probability distribution of different outcomes of a direct measurement on a quantum observable is described by the spectral projection-valued measure $\hat{P}(\cdot)$ on $(R, B(R))$ corresponding, due to the spectral theorem, to the self-adjoint operator representing this observable.

In the case of discrete spectrum of a measured quantum observable the famous von Neumann reduction postulate [1] prescribes the well-known "jump" of a state of a quantum system under a measurement. Specifically, if under a direct measurement upon a von Neumann observable

(1) $$\hat{B} = \sum_j \mathbf{l}_j \hat{P}_j ,$$

the initial state $\mathbf{r}_S$ of a quantum system is pure, that is, $\mathbf{r}_S = |\mathbf{y}_0\rangle\langle\mathbf{y}_0|$, and if under a single

measurement the outcome $l_j$ is observed, then in the frame of the von Neumann approach at the moment immediately after this measurement the quantum system "jumps" with certainty to the pure state

(2)
$$\frac{\hat{P}_j |\psi_0\rangle\langle\psi_0| \hat{P}_j}{\|\hat{P}_j\psi_0\|^2}.$$

The probability $m_j$ of the outcome $l_j$ is given by

(3)
$$m_j = \| P_j\psi_0 \|^2.$$

In the case of continuous spectrum of a measured quantum observable the description of a state change of a quantum system under a measurement is not formalized.

The simultaneous direct measurement of $n$ quantum observables is allowed if and only if the corresponding self-adjoint operators and, consequently, spectral projection-valued measures, commute. Such a measurement is described by the projection-valued measure

(4)
$$\hat{P}(E_1 \times E_2 \times ... \times E_n) = \hat{P}_1(E_1)\hat{P}_2(E_2)\cdot......\cdot\hat{P}_n(E_n)$$

on $(R^n, B(R^n))$ common for all $n$ commuting self-adjoint operators.

We would like to underline that in the case of discrete outcomes the original von Neumann approach gives both - the complete statistical description of a measurement and the complete stochastic description of the random behaviour of a quantum system under a single measurement (formula (2)).

The generalizations of von Neumann approach, to be discussed in the sequel, are caused by the fact that even for measurements with outcomes in $(R, B(R))$ this approach does not describe a state reduction of a quantum system in the general case where the spectrum of a measured quantum observable may be continuous or complicated, and it does not describe all measurements possible upon a quantum system.

*2.2. The description of a generalized quantum measurement*

In the further developments of quantum measurement theory [2-9] the mathematical notion of a probability operator-valued (POV) measure is used for the description of a probability distribution on a space of outcomes in the case of any measurement possible upon a quantum system.

Let $\Omega$ be a set of outcomes of the most general nature possible under a quantum measurement and $F$ be a $\sigma$-algebra of subsets of $\Omega$. Let $L(H_S)$ be the Banach space of all bounded linear operators on $H_S$.

A mapping $\hat{M}(\cdot): F \to L(H_S)$ is called a probability operator-valued measure, or a POV measure, for short, if $\hat{M}(\cdot)$ is a $\sigma$-additive measure on $(\Omega, F)$ with values $\hat{M}(E), E \in F$ that are positive bounded linear operators on $H_S$ such that the following condition is valid: $\hat{M}(\Omega) = \hat{I}$.

Given a POV measure, a scalar probability measure $m_{\rho_S}(\cdot)$ on $(\Omega, F)$, describing the probability distribution of possible outcomes of a measurement upon the quantum system, being at the instant before the measurement in the state $\hat{\rho}_S$, is given by

(5)
$$m_{\rho_S}(E) = \mathrm{tr}[\hat{\rho}_S \hat{M}(E)], \quad \forall E \in F.$$

In contrast to a spectral projection-valued measure on $(R, B(R))$, which is one-to-one defined by a self-adjoint operator, different possible measurements with oucomes in $(R, B(R))$, being described by different POV measures on $(R, B(R))$, may correspond to one and the same observable, represented by a self-adjoint operator.

A POV measure is sometimes called a generalized observable [3] or semiobservable [6] of a quantum system. A spectral projection-valued measure $\hat{P}(\cdot)$ on $(R, B(R))$ (and the corresponding self-adjoint operator, for short ) is called a von Neumann observable.

The notion of a POV measure does not, however, describe in any way a state change of a quantum system under a generalized quantum measurement. Thus, with respect to a quantum system it does not give the complete statistical description of a generalized quantum measurement.

## 2.3. Operational approach

The complete statistical description of any generalized quantum measurement is specified in the frame of the operational approach where the mathematical notion of a quantum instrument [2-6] plays a central role.

Specifically, a mapping $\hat{T}(\cdot)[\cdot]: F \times L(H_S) \to L(H_S)$ is called a quantum instrument if $\hat{T}(\cdot)$ is a $\sigma$-additive measure on $(\Omega, F)$ with values $\hat{T}(E), E \in F$, that are normal completely positive bounded linear maps $L(H_S) \to L(H_S)$ such that the following normality relation is valid: $\hat{T}(\Omega)[\hat{I}] = \hat{I}$.

From now on we shall only consider quantum instruments and henceforth we therefore suppress the term "quantum".

Given the instrument of a measurement, the POV measure of that measurement. is defined as

(6a) $$\hat{M}(E) = \hat{T}(E)[\hat{I}], \forall E \in F.$$

The scalar probability measure on $(\Omega, F)$, defining a probability distribution of possible outcomes under a measurement upon a quantum system being before the measurement in the state $\hat{\rho}_S$, is

(6b) $$\mu_{\hat{\rho}_S}(E) = \text{tr}[\hat{\rho}_S \hat{T}(E)[\hat{I}]].$$

The conditional expectation of any von Neumann observable $\hat{Z}$ at the instant immediately after the measurement, under the condition that the observed outcome belongs to the subset $E$, is given by

(7a) $$\text{Ex}\{\hat{Z} \mid E\} = \frac{\text{tr}[\hat{\rho}_S \hat{T}(E)[\hat{Z}]]}{\mu_{\hat{\rho}_S}(E)},$$

and the quantum mean value is

(7b) $$<\hat{Z}> \equiv \text{Ex}\{\hat{Z} \mid \Omega\} = \text{tr}[\hat{\rho}_S \hat{T}(\Omega)[\hat{Z}]].$$

The knowledge of an instrument gives the statistical description of a state change of a quantum system caused by a measurement [6]. The posterior (conditional) state (or density, or statistical, operator) of a quantum system $\hat{\rho}(E, \hat{\rho}_S)$, conditioned by the outcome being in $E$, is defined by the relation

(8a) $$\text{Ex}\{\hat{Z} \mid E\} = \frac{\text{tr}[\hat{\rho}_S \hat{T}(E)[\hat{Z}]]}{\mu_{\hat{\rho}_S}(E)} = \text{tr}[\hat{\rho}(E, \hat{\rho}_S) \hat{Z}].$$

The unconditional (prior) state $\hat{\rho}(\Omega, \hat{\rho}_S)$ of a quantum system defines the quantum mean value

(8b) $$<\hat{Z}> = \text{tr}[\hat{\rho}(\Omega, \hat{\rho}_S) \hat{Z}]$$

of a von Neumann observable $\hat{Z}$ at the instant after a measurement if the results of a measurement are ignored.

Any conditional state change of a quantum system can be completely described in the Hilbert space $H_S$ by a family of statistical operators $\{\hat{\rho}(\omega, \hat{\rho}_S), \omega \in \Omega\}$ called usually a family of posterior states [7,8]. For any instrument and a premeasurement state $\hat{\rho}_S$ of a quantum system the family $\{\hat{\rho}(\omega, \hat{\rho}_S), \omega \in \Omega\}$ always exists and is defined uniquely, $\mu_{\hat{\rho}_S}(\cdot)$ - almost everywhere, by

(9a) $$\text{tr}[\hat{\rho}_S \hat{T}(E)[\hat{A}]] = \int_{\omega \in E} \text{tr}[\hat{\rho}(\omega, \hat{\rho}_S) \hat{A}] \mu_{\hat{\rho}_S}(d\omega),$$

for $\forall \hat{A} \in L(H_S)$, $\forall E \in F$. From (8) and (9a) it follows that the family $\{\hat{r}(w, \hat{r}_S), w \in \Omega\}$ determines the conditional expectation by

(9b) $$\text{Ex}\{\hat{Z} \mid E\} = \frac{\int_{w \in E} \text{tr}[\hat{r}(w, \hat{r}_S)\hat{Z}] m_{\hat{r}_S}(dw)}{m_{\hat{r}_S}(E)}$$

and

(9c) $$<\hat{Z}> = \int_\Omega \text{tr}[\hat{r}(w, \hat{r}_S)\hat{Z}] m_{\hat{r}_S}(dw).$$

The posterior statistical operator $\hat{r}(E, \hat{r}_S)$ of a quantum system, conditioned by the outcome $w \in E$, is presented through the family of posterior states $\{\hat{r}(w, \hat{r}_S), w \in \Omega\}$ as

(10) $$\hat{r}(E, \hat{r}_S) = \frac{\int_{w \in E} \hat{r}(w, \hat{r}_S) m_{\hat{r}_S}(dw)}{m_{\hat{r}_S}(E)}.$$

There is a one-to-one correspondence between a POV measure and a family of posterior statistical operators on the one side and an instrument on the other side [7,8]. Knowing a POV measure and a family of posterior states one can reconstruct the instrument.

*2.4. Statistical realizations of an instrument*

As well as in the von Neumann approach as in the operational approach the notion of a projection-valued measure on $(\Omega, F)$ plays a fundamental role.

Introduce the following notation. Let $\hat{s}$ be a statistical operator on a separable Hilbert space $K$ and $\hat{Q}$ be an operator belonging to $L(H_S \otimes K)$. There exists [6] a uniquely determined normal completely positive bounded linear map $\text{E}_{\hat{s}}: L(H_S \otimes K) \to L(\hat{H}_S)$ such that the relation

(11) $$tr[\hat{r} \text{E}_{\hat{s}}[\hat{Q}]] = tr[(\hat{r} \otimes \hat{s})\hat{Q}]$$

is valid for any statistical operator $\hat{r}$ on $H_S$.

In [6] it was shown that for any instrument on a Borel space $(\Omega, F)$ with the values in $L(H_S)$ there exist a Hilbert space $K$, a statistical operator $\hat{s}$ on $K$, a unitary operator $\hat{U}$ and a projection-valued measure $\hat{I} \otimes \hat{P}(\cdot)$ on $H_S \otimes K$, such that the instrument can be presented in the form:

(12) $$\hat{T}(E)[\hat{A}] = \text{E}_{\hat{s}}[\hat{U}^+(\hat{A} \otimes \hat{P}(E))\hat{U}],$$

for $\forall E \in F$, $\forall \hat{A} \in L(H_S)$.

A 4-tuple

(13) $$\{K, \hat{s}, \hat{P}(\cdot), \hat{U}\}$$

is called a measuring process of the corresponding generalized observable (a POV measure) or a statistical realization of an instrument. For a given instrument a statistical realization always exists but may not be unique.

If in (13) the Hilbert space $K$ is separable then the corresponding statistical realization is called separable.

In quantum theory a Hilbert space $H_S$ of a system is always separable, while the value space is mostly a standard Borel space (that is a Borel space which is Borel isomorphic to a complete separable metric space).

If $(\Omega, F_B)$ is a standard Borel space and the Hilbert space $H_S$ of a quantum system is separable, then there exists a separable statistical realization of any instrument $\hat{T}(\cdot)[\cdot]$ on $(\Omega, F_B)$ [6].

In [1, p.442] von Neumann showed that the state reduction, first postulated by him in his projection postulate, can be formally derived in the scheme of a measuring process.

Consider [6] a von Neumann measuring process of the observable (1), which, with respect to the considered quantum system, results:

a) in the POV measure

$$\hat{M}(E) \equiv \hat{P}(E) = \sum_{l_j \in E} \hat{P}_j , \tag{14a}$$

on $(R, B(R))$, being the spectral projection-valued measure, corresponding to (1);

b) in the family of posterior states

$$\hat{r}(\{l_j\}, \hat{r}_S) = \frac{\hat{P}_j \hat{r}_S \hat{P}_j}{tr[\hat{r}_S \hat{P}_j]}, \tag{14b}$$

corresponding to von Neumann reduction postulate.

The unique instrument, corresponding to (14a,b), has the form

$$\hat{T}(E)[\hat{A}] = \sum_{l_j \in E} \hat{P}_j \hat{A} \hat{P}_j , \tag{14c}$$

for $\forall E \in B(R)$, $\forall \hat{A} \in L(H_S)$.

Let $\{\mathbf{y}_{jk}\}$ be the complete orthonormal set of eigenvectors of the observable (1):

$$\hat{B} \mathbf{y}_{jk} = l_j \mathbf{y}_{jk}, \quad \hat{P}_j = \sum_k |\mathbf{y}_{jk}\rangle\langle \mathbf{y}_{jk}| . \tag{15}$$

Let $K$ be another complex separable Hilbert space, $\{\mathbf{h}_i\}$ and $\mathbf{h}$ be, respectively, a complete set of orthonormal vectors and an unit vector in $K$. Let $\hat{U}$ be a unitary operator on $H_S \otimes K$, satisfying the relation

$$\hat{U}(\mathbf{y}_{jk} \otimes \mathbf{h}) = \mathbf{y}_{jk} \otimes \mathbf{h}_j . \tag{16}$$

The 4-tuple

$$\{K, |\mathbf{h}\rangle\langle\mathbf{h}|, \sum_{l_j \in E} |\mathbf{h}_j\rangle\langle \mathbf{h}_j|, \hat{U}\} \tag{17}$$

presents a von Neumann measuring process for the observable (1) or a separable statistical realization of the instrument (14c):

$$\hat{T}(E)[\hat{A}] = E_{|\mathbf{h}\rangle\langle\mathbf{h}|}[\hat{U}^+(\hat{A} \otimes \sum_{l_j \in E}|\mathbf{h}_j\rangle\langle\mathbf{h}_j|)\hat{U}] = \sum_{l_j \in E} \hat{P}_j \hat{A} \hat{P}_j . \tag{18}$$

We would like to emphasize that (17) presents a von Neumann measuring process of the observable (1) for any pair - a set $\{\mathbf{h}_i\}$ of orthonormal vectors and an unit vector $\mathbf{h}$ in $K$.

Thus, the concept of the direct measurement of the observable (1) in the frame of von Neumann approach corresponds to the description of different measuring processes, given by (17). We discuss this point in detail in section 4.

*2.5. Integral representations of an instrument*

In [10-12] it was proved (although, the contents of the corresponding theorems in [10,12] is slightly different) that for any instrument on $(\Omega, F_B)$ there exist a positive scalar measure $m(\cdot)$ on $(\Omega, F_B)$, a dense domain $D \subset H_S$, a countable family of functions $\mathbf{w} \to \hat{X}_k(\mathbf{w})$, defined for $m$-almost all $\mathbf{w}$, such that $\hat{X}_k(\mathbf{w})$ are linear operators from $D$ to $H_S$, satisfying the relation

$$\int_\Omega \sum_k \| \hat{X}_k(\mathbf{w}) \mathbf{y} \|^2 m(d\mathbf{w}) = \|\mathbf{y}\|^2, \quad \forall \mathbf{y} \in D \tag{19}$$

and

$$\langle \mathbf{y}, \hat{T}(E)[\hat{A}] \mathbf{y}\rangle = \int_{\mathbf{w} \in E} (\sum_k \langle \hat{X}_k(\mathbf{w}) \mathbf{y}, \hat{A} \hat{X}_k(\mathbf{w}) \mathbf{y}\rangle) m(d\mathbf{w}), \quad \forall \mathbf{y} \in D . \tag{20}$$

The representation (20) is similar to the Stinespring-Kraus representation for completely positive maps but according to [12] the operators $\hat{X}_k(\boldsymbol{w})$, involved in (20), are defined only on $D \subset H_S$.

If in (20) $k=1$, then such an instrument is called pure in [12].

From (20) and (9a) it follows that, if prior to the measurement the quantum system is in a pure state $\hat{r}_0 = |\boldsymbol{y}\rangle\langle\boldsymbol{y}|$ and $\boldsymbol{y} \in D$ then the family of posterior states $\{\hat{r}(\boldsymbol{w}, \hat{r}_0), \boldsymbol{w} \in \Omega\}$, describing the conditional state change of the quantum system, is given by

$$(21) \quad \hat{r}(\boldsymbol{w}, \hat{r}_0) = \frac{\sum_k |\hat{X}_k(\boldsymbol{w})\boldsymbol{y}\rangle\langle\hat{X}_k(\boldsymbol{w})\boldsymbol{y}|}{\sum_k \|\hat{X}_k(\boldsymbol{w})\boldsymbol{y}\|^2}.$$

For the case of continuous in time nondemolition observation of an open system the representations of an instrument, similar to (20), were considered in [13,14.] (cf. also references there).and in [15-17].

In the physical literature on quantum measurements, in the special case when $\Omega = R$ and the spectrum of the measured quantum quantity is discrete, the formulae for the POV measure and the posterior states, similar to (20), (21), were presented in [18,19].

### 3. Quantum stochastic representation of an instrument

The aim of the present section is to analyse if there is any mathematical background for the description of a generalised quantum measurement via probability scalar measures on $(\Omega, F_B)$ and operator-valued functions, defined with respect to these measures and describing in a Hilbert space $H_S$ the conditional behaviour of a quantum system under a measurement.

It was noted in section 2.4 that for a given instrument a statistical realization (a measuring process) always exists but may not be unique. This mathematical fact corresponds to a clear physical situation when for different quantum measurements their statistical description may be the same.

An integral representation of an instrument also has a clear physical interpretation since it allows, in principle, to consider not only a statistical description of a generalized quantum measurement but also the stochastic conditional evolution of the quantum system under a single generalized measurement.

However, the integral representation of an instrument (20) is not based on any invariants of the corresponding measuring processes.

That is why, we are now interested what is the most general form of an integral representation of an instrument and what is the correspondence, in the most general case, between classes of statistical realizations and classes of integral representations of the same instrument.

In quantum theory any physically based problem must be formulated in unitarily equivalent terms and the results of its consideration must not be dependent neither on the choice of a special representation picture (Schrödinger, Heisenberg or interaction) nor on the choice of a basis in the Hilbert space. That is why, in section 3.1 we introduce the notion of the class of unitarily equivalent separable statistical realizations of an instrument and consider its invariants.

We find (section 3.2) the most general form (theorem 1) of integral representation of an instrument. This form differs from the integral representation (20). The most important difference is due to the orthogonality relation, which is not present in integral representations of an instrument, available in the mathematical and physical literature [10-13,18,19]. We prove in the most general case that in the case of a finite positive scalar measure ( $\boldsymbol{m}(\cdot)$ in the notation of (19)) the integral representation of an instrument is given through the $\boldsymbol{m}$-measurable operator-valued functions on all of $H_S$.

We introduce the notion of a stochastic realization of an instrument (section 3.3). We show that for any instrument there exists a stochastic realization (proposition 1) and establish a one-to-one correspondence between the class of unitarily equivalent separable statistical realizations and the equivalence class of stochastic realizations of the instrument (theorems 2, 3).

We call a stochastic realization of an instrument quantum stochastic (section 3.4) if it has the factorized form (52a). We show that for any instrument there exists a quantum stochastic realization (proposition 2). We further single out the invariant class of unitarily equivalent separable statistical realizations for which the corresponding equivalence class of stochastic realizations contains only invariant quantum stochastic realizations, specified by (53a). We prove (theorem 4) that for the invariant class of unitarily equivalent statistical realizations the integral representation of an instrument is the same for all statistical realizations from this class and is wholly determined by the invariants of this class. We call the special form (56) of the integral representation of an instrument, corresponding to an invariant class of unitarily and phase equivalent statistical realizations, a quantum stochastic representation of an instrument. The term "quantum" reflects the importance of this kind of integral representation of an instrument for the description of generalized direct quantum measurements (cf. section 4).

The general mathematical results, derived in this section, can be used for the quantum measurement theory and the theory of open systems. These results allow us to introduce in section 4 a new general approach, the quantum stochastic approach (QSA), to the description of generalized direct quantum measurements.

From now on we shall only consider complex separable Hilbert spaces and separable statistical realizations and henceforth we therefore suppress the term"separable".

*3.1. Unitary invariants of a statistical realization*

Let

(22a) $$\hat{T}(E)[\hat{A}]$$

be an instrument on a standard Borel space $(\Omega, F_B)$ with values in $L(H_S)$, where $H_S$ is a complex Hilbert space of the quantum system.

Let $\boldsymbol{g} = \{K, \hat{\boldsymbol{s}}, \hat{P}(\cdot), \hat{U}\}$ be a statistical realization of the instrument (22a), that is,

(22b) $$\hat{T}(E)[\hat{A}] = \mathrm{E}_{\boldsymbol{s}}[\hat{U}^+(\hat{A} \otimes \hat{P}(E))\hat{U}].$$

Consider some general properties of $\mathrm{E}_{\boldsymbol{s}}[\hat{Q}]$, defined by (11), where $\hat{Q}$ is an operator belonging to $L(H_S \otimes K)$.

Let $\hat{\boldsymbol{s}} = \sum_i \boldsymbol{l}_i \hat{p}_i$, $\hat{p}_i \hat{p}_j = \boldsymbol{d}_{ij} \hat{p}_j$ be the spectral decomposition of the statistical operator $\hat{\boldsymbol{s}}$ and $k(\boldsymbol{l}_i)$ be the multiplicity (which is always finite) of the positive eigenvalue $\boldsymbol{l}_i$. Then, letting $\hat{\boldsymbol{s}}_i = (k(\boldsymbol{l}_i))^{-1} \hat{p}_i$, we have

(23a) $$\mathrm{E}_{\boldsymbol{s}}[\hat{Q}] = \sum_i \boldsymbol{l}_i k(\boldsymbol{l}_i) \mathrm{E}_{\boldsymbol{s}_i}[\hat{Q}],$$

where the sum is convergent [6] in the weak operator topology.

Under a unitary transform $\hat{W} : K' \to K$, we have the following relation

(23b) $$\mathrm{E}_{\boldsymbol{s}}[\hat{Q}] = \mathrm{E}_{\boldsymbol{s}'}[\hat{Q}']$$

with the statistical operator $\hat{\boldsymbol{s}}'$ on $K'$ and the operator $\hat{Q}' \in L(H_S \otimes K')$ being given by

(23c) $$\hat{\boldsymbol{s}}' = \hat{W}^{-1} \hat{\boldsymbol{s}} \hat{W}, \quad \hat{Q}' = (\hat{I} \otimes \hat{W}^{-1}) \hat{Q} (\hat{I} \otimes \hat{W}) .$$

Consider a statistical realization $\boldsymbol{g} = \{K, \hat{\boldsymbol{s}}, \hat{P}(\cdot), \hat{U}\}$ of an instrument (22a).

**Definition.** *We shall say that a statistical realization $\boldsymbol{g}' = \{K', \hat{\boldsymbol{s}}', \hat{P}'(\cdot), \hat{U}'\}$ is unitarily equivalent to $\boldsymbol{g}$ if there exists a unitary transform $\hat{W} : K' \to K$ under which:*

(24a) $$\hat{\boldsymbol{s}}' = \hat{W}^{-1} \hat{\boldsymbol{s}} \hat{W}, \; \hat{P}'(\cdot) = \hat{W}^{-1} \hat{P}(\cdot) \hat{W}, \; \hat{U}' = (\hat{I} \otimes \hat{W}^{-1}) \hat{U} (\hat{I} \otimes \hat{W}).$$

**Definition.** *All properties of a statistical realization, which do not change under a unitary transform (24a), we shall call unitary invariants of the statistical realization.*

Due to (22b), (23b) and (23c) the instrument is a unitary invariant of the statistical realization.

We shall also say that a statistical realization $\boldsymbol{g}_x = \{K, \hat{\boldsymbol{S}}, \hat{P}(\cdot), e^{ix}\hat{U}\}$ is phase-equivalent to the statistical realization $\boldsymbol{g}$. Let $G(\boldsymbol{g}_x)$ be the set of all separable statistical realizations of an instrument (22) unitarily equivalent to a statistical realization $\boldsymbol{g}_x$.

Introduce $G_g = \{G(\boldsymbol{g}_x), x \in R\}$ - the class of all statistical realizations unitarily and phase equivalent to the statistical realization $\boldsymbol{g}$. The class $G_g$ includes, in particular, all unitarily and phase equivalent statistical realizations corresponding to one and the same Hilbert space $K$.

**Definition.** *We shall say that some property is an invariant of a class of statistical realizations if this property is the same for all statistical realizations from this class.*

The dimension $D_g$ of Hilbert spaces $K$ in statistical realizations from the class $G_g$ is the simplest invariant of this class.

In general, the same instrument induces different classes $G_g$, but all these classes have a common invariant - the instrument itself.

Let

(24b)
$$\{H_R, \hat{\boldsymbol{S}}_R, \hat{P}_R(\cdot), \hat{U}_R\}$$

be any statistical realization from the class $G_g$ on some fixed Hilbert space $H_R$. Consider on $(\Omega, F_B)$ the family of positive scalar Borel measures $\{\boldsymbol{m}_j(\cdot) = \langle \boldsymbol{j}, \hat{P}_R(\cdot)\boldsymbol{j}\rangle, \forall \boldsymbol{j} \in H_R\}$, induced by the projection-valued measure $\hat{P}_R(\cdot)$. For any projection-valued measure in the Hilbert space $H_R$ there exists [20] $\tilde{\boldsymbol{j}} \in H_R$ such that with respect to a subset $E \in F_B$ the equations $\boldsymbol{m}_{\tilde{j}}(E) = 0$ and $\hat{P}_R(E) = \hat{0}$ are equivalent. The element $\tilde{\boldsymbol{j}} \in H_R$ is said to be an element of maximum type [20] for the projection-valued measure $\hat{P}_R(\cdot)$. Denote by $[\boldsymbol{m}_j]$ the type of the scalar measure $\boldsymbol{m}_j(\cdot)$ ( i.e. $[\boldsymbol{m}_j]$ is the class of positive scalar measures equivalent to $\boldsymbol{m}_j(\cdot)$).

**Definition.** *The spectral type $[\hat{P}_R(\cdot)]$ of a projection-valued measure $\hat{P}_R(\cdot)$ on $(\Omega, F_B)$ is defined to be equal to the type $[\boldsymbol{m}_{\tilde{j}}]$ of the positive scalar Borel measure $\boldsymbol{m}_{\tilde{j}}(\cdot) = \langle \tilde{\boldsymbol{j}}, \hat{P}_R(\cdot)\tilde{\boldsymbol{j}}\rangle$, induced by an element $\tilde{\boldsymbol{j}} \in H_R$ of the maximum type [20].*

Let $\boldsymbol{n}(\cdot)$ be a positive scalar Borel measure on $(\Omega, F_B)$ of the type $[\boldsymbol{n}(\cdot)] = [\hat{P}_R(\cdot)]$. For any $\boldsymbol{f} \in H_R$ introduce the subset

(25a)
$$\Omega(\boldsymbol{f}) = \{\boldsymbol{w} \mid \boldsymbol{w} \in \Omega, \frac{d\boldsymbol{m}_f}{d\boldsymbol{n}}(\boldsymbol{w}) > 0\},$$

which is defined $\boldsymbol{n}$- almost everywhere ($\boldsymbol{n}$-a.e.) and does not depend on the choice of the scalar measure $\boldsymbol{n}(\cdot)$ on $(\Omega, F_B)$ out of the class of equivalent scalar measures of the type $[\hat{P}_R(\cdot)]$.

The following statements are valid $\boldsymbol{n}$-a.e. [20]:

(25b) $\qquad \Omega(\boldsymbol{f}) \subset \Omega(\boldsymbol{j}) \Leftrightarrow [\boldsymbol{m}_f] \prec [\boldsymbol{m}_j]$ ;

(25c) $\qquad \Omega(\boldsymbol{f}) = \Omega(\boldsymbol{j}) \Leftrightarrow [\boldsymbol{m}_f] = [\boldsymbol{m}_j]$ ;

(25d) $\qquad \hat{P}_R(\Omega(\boldsymbol{f}))\boldsymbol{f} = \boldsymbol{f}, \quad \forall \boldsymbol{f} \in H_R$ ;

(25e) $\quad\quad\quad\quad\quad\quad\quad\quad \Omega(\boldsymbol{h}_1) \cap \Omega(\boldsymbol{h}_2) = \varnothing \;\Rightarrow\; <\boldsymbol{h}_1, \hat{P}_R(\cdot)\boldsymbol{h}_2> = 0;$

(25f) $\quad\quad\quad\quad\quad\quad\quad\quad <\boldsymbol{h}_1, \hat{P}_R(\cdot)\boldsymbol{h}_2> = 0 \;\Rightarrow\; \Omega(\boldsymbol{h}_1 + \boldsymbol{h}_2) = \Omega(\boldsymbol{h}_1) \cup \Omega(\boldsymbol{h}_2);$

If

(25g) $\quad\quad\quad\quad\quad\quad\quad\quad [\boldsymbol{m}_f] = [\hat{P}_R(\cdot)], \;\; \boldsymbol{h} = \hat{P}_R(E)\boldsymbol{f}, \; E \in F_B,$

then

(25h) $\quad\quad\quad\quad\quad\quad\quad\quad \Omega(\boldsymbol{h}) = E.$

For any projection-valued measure $\hat{P}_R(\cdot)$ on $(\Omega, F_B)$ there exists [20] a family of elements $\{\boldsymbol{h}_j, \boldsymbol{h}_j \in H_R, \; j = 1,...,m; \; 1 \le m \le \infty\}$, satisfying $<\boldsymbol{h}_l, \hat{P}_R(\cdot)\boldsymbol{h}_k> = 0, \;\; \forall l \ne k$, such that

(26a) $\quad\quad\quad\quad\quad\quad\quad\quad H_R = \sum_j \oplus H_{\boldsymbol{h}_j},$

$$H_{\boldsymbol{h}_j} = \{\hat{Z}_f \boldsymbol{h}_j \mid f \in S(\Omega, \boldsymbol{n}), \boldsymbol{h}_j \in D_f\}$$

and

(26b) $\quad\quad\quad\quad\quad\quad\quad\quad [\hat{P}_R] = [\boldsymbol{m}_{\boldsymbol{h}_1}] \succ [\boldsymbol{m}_{\boldsymbol{h}_2}] \succ ....$

In (26a) $S(\Omega, \boldsymbol{n})$ is the class of $\boldsymbol{n}$-measurable, $\boldsymbol{n}$-a.e. finite functions: $\Omega \to C$; $\hat{Z}_f$ is the operator defined by the relation

(26c) $\quad\quad\quad\quad\quad\quad\quad\quad \hat{Z}_f = \int_\Omega f(\boldsymbol{w}) \hat{P}_R(d\boldsymbol{w})$

with the domain $D_f = \{\boldsymbol{y} \in H_R \mid \int_\Omega |f(\boldsymbol{w})|^2 \boldsymbol{m}_y(d\boldsymbol{w}) < \infty\}$.

If $m > 1$, then the decomposition (26a) of the Hilbert space $H_R$ is not unique.

From (25) and (26b) it follows that

(27a) $\quad\quad\quad\quad\quad\quad\quad\quad \Omega = \Omega(\boldsymbol{h}_1) \supset \Omega(\boldsymbol{h}_2) \supset ....$

Introduce the sets $\Omega_k, \; k = 1,...,m$ by the relations

(27b) $\quad\quad\quad\quad\quad\quad\quad\quad \Omega_k = \Omega(\boldsymbol{h}_k) \setminus \Omega(\boldsymbol{h}_{k+1}), \;\; k < m,$

$$\Omega_m = \bigcap_{k=1,...,m} \Omega(\boldsymbol{h}_k).$$

**Definition.** *The $\hat{P}_R$- measurable function $N_{P_R} : \Omega \to \{1,2,.....n,....\}$ defined $\hat{P}_R$- almost everywhere by the relation*

(27c) $\quad\quad\quad\quad\quad\quad\quad\quad N_{P_R}(\boldsymbol{w}) = k, \;\; for \;\; \boldsymbol{w} \in \Omega_k, \;\; k = 1,...,m$

*is called a multiplicity function of the projection-valued measure $\hat{P}_R(\cdot)$ on $(\Omega, F_B)$* [20].

The type $[\hat{P}_R(\cdot)]$ and the multiplicity function $N_{P_R}(\boldsymbol{w})$ characterize the projection-valued measure $\hat{P}_R(\cdot)$ on $(\Omega, F_B)$ up to unitary equivalence [20] (see also the formula (28e) below). Since $[\hat{P}_R(\cdot)]$ and $N_{P_R}(\boldsymbol{w})$ are unitary invariants of $\hat{P}_R(\cdot)$, they are unitary invariants of any statistical realization from the class $G_g$ and they are invariants of the class $G_g$.

Let

(28a) $\quad\quad\quad\quad\quad\quad\quad\quad H(\boldsymbol{n}, N; Y) = \int_\Omega \oplus_Y H(\boldsymbol{w}) \boldsymbol{n}(d\boldsymbol{w})$

be the direct integral [20-22] of Hilbert spaces $H(\boldsymbol{w})$ on $(\Omega, F_B)$, induced by:

- a positive scalar Borel measure $\boldsymbol{n}(\cdot)$ of the type $[\boldsymbol{n}(\cdot)] = [\hat{P}_R(\cdot)]$;

- the dimension function $N(\boldsymbol{w}) = \dim H(\boldsymbol{w})$, being equal $\boldsymbol{n}$-a.e. on $\Omega$ to the multiplicity function $N_{P_R}(\boldsymbol{w})$ of the projection-valued measure $\hat{P}_R(\cdot)$ on $(\Omega, F_B)$;
- an orthonormal base of measurability $Y = \{e_n\}$, $n = 1,...,l$, where the positive integer $l$ is equal to $\boldsymbol{n}$-$\sup\{N(\boldsymbol{w}), \boldsymbol{w} \in \Omega\}$, which is defined as $\inf\{c \in R : N(\boldsymbol{w}) \leq c, \boldsymbol{n}\text{-a.e.}\}$.

For any $\boldsymbol{w} \in \Omega$ the set $\{e_n(\boldsymbol{w}), n = 1,...,N(\boldsymbol{w})\}$ represents an orthonormal basis in the Hilbert space $H(\boldsymbol{w})$. Every measurable function $e_n(\boldsymbol{w})$ is defined to be $e_n(\boldsymbol{w}) \equiv 0$ for $n > N(\boldsymbol{w})$. Notice that $\|e_1(\boldsymbol{w})\|_{H(\boldsymbol{w})} = 1$, $\boldsymbol{n}$-a.e. on $\Omega$.

Recall [cf.20] that the scalar product in the separable Hilbert space $H(\boldsymbol{n}, N; Y)$ is defined by

(28b) $$<f, g> = \int_\Omega <f(\boldsymbol{w}), g(\boldsymbol{w})>_{H(\boldsymbol{w})} \boldsymbol{n}(d\boldsymbol{w}),$$

where the function

(28c) $$<f(\boldsymbol{w}), g(\boldsymbol{w})>_{H(\boldsymbol{w})} = \sum_{n=1}^{N(\boldsymbol{w})} <f(\boldsymbol{w}), e_n(\boldsymbol{w})>_{H(\boldsymbol{w})} <e_n(\boldsymbol{w}), g(\boldsymbol{w})>_{H(\boldsymbol{w})}$$

is $\boldsymbol{n}$-measurable since on the right hand side in (28c) we have a convergent series of measurable functions.

Two direct integrals $H(\boldsymbol{n}, N; Y)$ and $H'(\boldsymbol{n}', N'; Y')$, induced by equivalent measures $\boldsymbol{n} \sim \boldsymbol{n}'$ and equal, $\boldsymbol{n}$-a.e. on $\Omega$, dimension functions $N(\boldsymbol{w}) = N'(\boldsymbol{w})$, but possibly different orthonormal bases of measurability $Y$ and $Y'$, are isometrically isomorphic to each other.

Since we are interested in finding unitary invariants of a statistical realization, we may take any of the equivalent measures. We may also take any orthonormal base of measurability $Y$.

We take a finite measure $\boldsymbol{n}(\Omega) < \infty$, since in this case any element of the orthonormal base of measurability $Y = \{e_n\}$ belongs to $H(\boldsymbol{n}, N; Y)$.

Let $\hat{X}_{\boldsymbol{n}}(\cdot)$ be the projection-valued measure on $H(\boldsymbol{n}, N; Y)$, defined by the relation

(28d) $$(\hat{X}_{\boldsymbol{n}}(E)g)(\boldsymbol{w}) = \boldsymbol{c}_E(\boldsymbol{w})g(\boldsymbol{w}), \quad \forall g \in H(\boldsymbol{n}, N; Y),$$

$\boldsymbol{n}$-a.e. on $\Omega$, where $\boldsymbol{c}_E(\cdot)$ is the characteristic function of a subset $E \in F_B$.

Then there exists [20] a unitary transform $\hat{R} : H_R \to H(\boldsymbol{n}, N; Y)$ such that

(28e) $$\hat{P}_R(E) = \hat{R}^{-1} \hat{X}_{\boldsymbol{n}}(E) R.$$

The spectral type $[\hat{X}_{\boldsymbol{n}}(\cdot)]$ is equal to $[\hat{P}_R(\cdot)] = [\boldsymbol{n}(\cdot)]$.

The statistical realization

(28f) $$\{H(\boldsymbol{n}, N; Y), \hat{\boldsymbol{S}}_{\boldsymbol{n}}, \hat{X}_{\boldsymbol{n}}(\cdot), \hat{U}_{\boldsymbol{n}}\},$$
$$\hat{\boldsymbol{S}}_{\boldsymbol{n}} = \hat{R}\hat{\boldsymbol{S}}_R \hat{R}^{-1}, \quad \hat{U}_{\boldsymbol{n}} = (\hat{I} \otimes \hat{R})\hat{U}_R(\hat{I} \otimes \hat{R}^{-1})$$

is unitarily equivalent to the statistical realization (24b) and belongs to the class $G_g$.

Any other statistical realization $\{H(\boldsymbol{n}, N; Y), \hat{\boldsymbol{S}}'_{\boldsymbol{n}}, \hat{X}_{\boldsymbol{n}}(\cdot), \hat{U}'_{\boldsymbol{n}}\}$ from the class $G_g$ on $H(\boldsymbol{n}, N; Y)$ must be unitarily equivalent to (28f) with a unitary transform $\hat{Z} : H(\boldsymbol{n}, N; Y) \to H(\boldsymbol{n}, N; Y)$, commuting with the projection-valued measure $\hat{X}_{\boldsymbol{n}}(\cdot)$ and, consequently, being a decomposable operator on $H(\boldsymbol{n}, N; Y)$ [20]. As any decomposable operator on $H(\boldsymbol{n}, N; Y)$, the unitary operator $\hat{Z}$ is presented by the relation:

(28g) $$(\hat{Z}g)(\boldsymbol{w}) = \hat{z}(\boldsymbol{w})g(\boldsymbol{w}), \quad \hat{z}(\boldsymbol{w}) \in L(H(\boldsymbol{w})), \quad \hat{z}^+(\boldsymbol{w})\hat{z}(\boldsymbol{w}) = \hat{z}(\boldsymbol{w})\hat{z}^+(\boldsymbol{w}) = \hat{I}_{H(\boldsymbol{w})},$$
$$\forall g \in H(\boldsymbol{n}, N; Y),$$

$\boldsymbol{n}$-a.e. on $\Omega$.

Due to (23b) the instrument, being a unitary invariant of the statistical realization $\boldsymbol{g}$ and an invariant of the class $G_g$, is given through the elements of the considered statistical realisation (28f) as

(29a) $$\hat{T}(E)[\hat{A}] = \mathrm{E}_{\boldsymbol{s_n}}[\hat{U}_n^+(\hat{A} \otimes \hat{X}_n(E))\hat{U}_n],$$

for $\forall E \in F_B, \quad \forall \hat{A} \in L(H_S)$.

Consider the spectral decomposition of the statistical operator $\hat{\boldsymbol{s}}_n$:

(29b) $$\hat{\boldsymbol{s}}_n = \sum_{i=1}^{N_g} \boldsymbol{a}_g^{(i)} \hat{p}_i, \quad \hat{p}_i \hat{p}_j = \boldsymbol{d}_{ij} \hat{p}_i,$$

$$\boldsymbol{a}_g^{(i)} > 0, \quad \sum_{i=1}^{N_g} \boldsymbol{a}_g^{(i)} k_g(\boldsymbol{a}_g^{(i)}) = 1.$$

The set $\boldsymbol{a}_g = \{\boldsymbol{a}_g^{(i)}, i=1,...,N_g\}$ of different positive eigenvalues of $\hat{\boldsymbol{s}}_n$, the positive integer $N_g$ (it may be infinite) and the multiplicities $k_g(\boldsymbol{a}_g^{(i)}) < \infty$ of positive eigenvalues $\boldsymbol{a}_g^{(i)}$ are unitary invariants of the statistical realization (28f) and they are invariants of the class $G_g$. Let $\{\boldsymbol{j}_{ik}\}$ be the complete orthonormal set of eigenvectors of the statistical operator $\hat{\boldsymbol{s}}_n$, then in (29b) we have

(29c) $$\hat{p}_i = \sum_{k=1}^{k_g(\boldsymbol{a}_g^{(i)})} |\boldsymbol{j}_{ik}><\boldsymbol{j}_{ik}|.$$

For any index $i$, for which the multiplicity $k_g(\boldsymbol{a}_g^{(i)}) > 1$, the set of eigenvectors $\{\boldsymbol{j}_{ik}, k=1,...k_g(\boldsymbol{a}_g^{(i)})\}$ is defined uniquely up to unitary equivalence, corresponding to different choices of the basis in the subspace $\hat{p}_i H(\boldsymbol{n}, N; Y)$.

Thus, $[\hat{P}_R], N_{P_R}, \boldsymbol{a}_g, N_g, \{k_g(\boldsymbol{a}_g^{(i)})\}$ are unitary invariants of the statistical realization $\boldsymbol{g}$ and invariants of the class $G_g$.

From (23c) it follows that we can decompose the instrument (29a) in the following form:

(30a) $$\hat{T}(E)[\hat{A}] = \sum_{i=1}^{N_g} \boldsymbol{a}_g^{(i)} k_g(\boldsymbol{a}_g^{(i)}) \mathrm{E}_{\boldsymbol{s}_n^{(i)}}[\hat{U}_n^+(\hat{A} \otimes \hat{X}_n(E))\hat{U}_n],$$

for $\forall E \in F_B, \quad \forall \hat{A} \in L(H_S)$. In (30a) $\hat{\boldsymbol{s}}_n^{(i)}$ is the statistical operator defined by

(30b) $$\hat{\boldsymbol{s}}_n^{(i)} = (k_g(\boldsymbol{a}_g^{(i)}))^{-1} \hat{p}_i.$$

Denote by

(30c) $$\hat{T}_i(E)[\hat{A}] = \mathrm{E}_{\boldsymbol{s}_n^{(i)}}[\hat{U}_n^+(\hat{A} \otimes \hat{X}_n(E))\hat{U}_n]$$

any "$i$" instrument in the decomposition (30a).

Introduce an equivalence relation on the space $G = \{G_g, \boldsymbol{g} \in \Gamma\}$ of all the classes of unitarily and phase equivalent statistical realizations of the instrument (22a) in the following way.

Two elements $G_{g_1}$ and $G_{g_2}$ of $G$ are equivalent if for any statistical realization $\boldsymbol{g}_1$ from the class $G_{g_1}$ and any statistical realization $\boldsymbol{g}_2$ from the class $G_{g_2}$ there exists a unitary transform $\hat{W}: K_1 \to K_2$ and a real number $\boldsymbol{x}$ under which

$$\hat{P}^{(1)}(\cdot) = \hat{W}^{-1}\hat{P}^{(2)}(\cdot)\hat{W} \qquad \hat{U}^{(1)} = e^{i\boldsymbol{x}}(\hat{I} \otimes \hat{W}^{-1})\hat{U}^{(2)}(\hat{I} \otimes \hat{W})$$

(30d) $$\hat{\boldsymbol{s}}_{g_1} = \sum_{i=1}^{N_{g_1}} \boldsymbol{a}_{g_1}^{(i)} k_{g_1}(\boldsymbol{a}_{g_1}^{(i)}) \hat{\boldsymbol{s}}_{g_1}^{(i)} \qquad \hat{\boldsymbol{s}}_{g_2} = \sum_{i=1}^{N_{g_2}} \boldsymbol{a}_{g_2}^{(i)} k_{g_2}(\boldsymbol{a}_{g_2}^{(i)}) \boldsymbol{s}_{g_2}^{(i)}$$

$$N_{g_1} = N_{g_2} \qquad k_{g_1}(\boldsymbol{a}_{g_1}^{(i)}) = k_{g_2}(\boldsymbol{a}_{g_2}^{(i)}) \qquad \hat{\boldsymbol{s}}_{g_1}^{(i)} = \hat{W}^{-1}\boldsymbol{s}_{g_2}^{(i)}\hat{W} \qquad \forall i = 1,...,N_{g_1}.$$

We denote an equivalence class in $G$ by $[G_g]$ if contains the class $G_g$.

Due to (30d) and the property (23b,c) all instruments $\hat{T}_i(\cdot)[\cdot]$, $i=1,...,N_g$, defined by (30c), are invariants of the equivalence class $[G_g]$. Consequently, they are invariants of the class $G_g$. Then, it follows that (30a) is an invariant decomposition for the class $G_g$, that is the same for all statistical realizations from this class.

Introduce on $(\Omega, F_B)$ the probability scalar measures:

(31a)
$$\boldsymbol{n}_g(E) = tr[\hat{\boldsymbol{s}}_n \hat{X}_n(E)] = \sum_{i=1}^{N_g} \boldsymbol{a}_g^{(i)} k_g(\boldsymbol{a}_g^{(i)}) \boldsymbol{n}_g^{(i)}(E),$$

$$\boldsymbol{n}_g^{(i)}(E) = tr[\hat{\boldsymbol{s}}_n^{(i)} \hat{X}_n(E)], \quad i=1,...,N_g.$$

These probability scalar measures are invariants of the class $G_g$. The probability scalar measures $\boldsymbol{n}_g^{(i)}(\cdot)$, $i=1,...,N_g$ are also invariants of the equivalence class $[G_g]$.

From (31a) and (28b,d) it follows that the probability scalar measures $\boldsymbol{n}_g(\cdot), \boldsymbol{n}_g^{(i)}(\cdot)$ are absolutely continuous with respect to the positive scalar measure $\boldsymbol{n}(\cdot)$ in the direct integral $H(\boldsymbol{n}, N; Y)$:

(31b)
$$\boldsymbol{n}_g^{(i)}(E) = \int_{\boldsymbol{w} \in E} \boldsymbol{p}_i(\boldsymbol{w}) \boldsymbol{n}(d\boldsymbol{w}).$$

Substituting (30b) into (31a) and considering (28d) and (29c), we get the following expression for the density of the probability scalar measure $\boldsymbol{n}_g^{(i)}(\cdot)$ with respect to $\boldsymbol{n}(\cdot)$:

(31c)
$$\boldsymbol{p}_i(\boldsymbol{w}) = tr_{H(\boldsymbol{w})}[\hat{\boldsymbol{s}}_n^{(i)}(\boldsymbol{w})],$$

where we denoted

(31d)
$$\hat{\boldsymbol{s}}_n^{(i)}(\boldsymbol{w}) = (k_g(\boldsymbol{a}_g^{(i)}))^{-1} \sum_{k=1}^{k_g(\boldsymbol{a}_g^{(i)})} \boldsymbol{j}_{ik}(\boldsymbol{w}) < \boldsymbol{j}_{ik}(\boldsymbol{w}), \cdot >_{H(\boldsymbol{w})}.$$

The density $\boldsymbol{p}_i(\boldsymbol{w})$ does not depend on the choice of the basis $\{\boldsymbol{j}_{ik}, k=1,...,k_g(\boldsymbol{a}_g^{(i)})\}$ in the subspace $\hat{p}_i H(\boldsymbol{n}, N; Y)$.

Introduce also the operator-valued measures

(32a)
$$\hat{\Theta}_{\tilde{a}}(E) = \mathrm{E}_{\hat{\boldsymbol{s}}_n}[(\hat{I} \otimes \hat{X}_n(E))\hat{U}_n] = \sum_{i=1}^{N_g} \boldsymbol{a}_g^{(i)} k_g(\boldsymbol{a}_g^{(i)}) \Theta_{\tilde{a}}^{(i)}(E),$$

(32b)
$$\hat{\Theta}_{\tilde{a}}^{(i)}(E) = \mathrm{E}_{\hat{\boldsymbol{s}}_n^{(i)}}[(\hat{I} \otimes \hat{X}_n(E))\hat{U}_n]$$

on $(\Omega, F_B)$ with values in $L(H_S)$.

These operator-valued measures are invariants (up to phase equivalence) of the class $G_g$. The measures $\hat{\Theta}_{\tilde{a}}^{(i)}(E)$ are also invariants (up to phase equivalence) of the equivalence class $[G_g]$. Thus, we derived the following sets of invariants of the class $G_g$

(32c)
$$[\hat{P}_R], N_{P_R}, \boldsymbol{a}_g, N_g, \{k_g(\boldsymbol{a}_g^{(i)})\}, \{\boldsymbol{n}_g^{(i)}(\cdot)\}, \boldsymbol{n}_g(\cdot), \{\hat{\Theta}_g^{(i)}(\cdot)\}, \hat{\Theta}_g(\cdot)$$

and of the equivalence class $[G_g]$

(32d)
$$[\hat{P}_R], N_{P_R}, N_g, \{k_g(\boldsymbol{a}_g^{(i)})\}, \{\boldsymbol{n}_g^{(i)}(\cdot)\}, \{\hat{\Theta}_g^{(i)}(\cdot)\}.$$

*3.2. General form of the integral representation of an instrument*

As we have already mentioned the same instrument $\hat{T}(E)[\hat{A}]$ induces different classes of unitarily and phase equivalent statistical realizations, but all these classes have a common invariant - the instrument itself. Consider for the instrument $\hat{T}(E)[\hat{A}]$ and the instruments $\hat{T}_i(E)[\hat{A}]$, $i=1,...,N_g$, introduced by (30c), possible integral representations, corresponding to the definite class $G_g$. The most general form of integral representation of an instrument is specified in theorem1.

Take some fixed orthonormal basis $\{\boldsymbol{j}_{ik}, k=1,...,k_g(\boldsymbol{a}_g^{(i)})\}$ in the subspaces $\hat{p}_i H(\boldsymbol{n},N;Y)$. For any index $i=1,...,N_g$ the unitary transformation from one basis to another is described by:

$$\boldsymbol{j}_{ik} = \sum_{p=1}^{k_g(\boldsymbol{a}_g^{(i)})} \boldsymbol{J}_{kp}^{(i)} \tilde{\boldsymbol{j}}_{ip}, \tag{33a}$$

where $\{\boldsymbol{J}_{kp}^{(i)}\}$ is any unitary matrix with elements being complex numbers.

Introduce on the Hilbert space $H(\boldsymbol{n},N;Y)$ the decomposable projections $\hat{Q}_n$, $n=1,...,l$ (with $l$ being equal to $\boldsymbol{n}$-$\sup\{N(\boldsymbol{w}), \boldsymbol{w}\in\Omega\}$), which are defined by the relation

$$(\hat{Q}_n g)(\boldsymbol{w}) = \hat{q}_n(\boldsymbol{w}) g(\boldsymbol{w}), \quad \forall g \in H(\boldsymbol{n},N;Y), \tag{33b}$$
$$\hat{q}_n(\boldsymbol{w}) = e_n(\boldsymbol{w}) < e_n(\boldsymbol{w}), \cdot >_{H(\boldsymbol{w})},$$

$\boldsymbol{n}$-a. e. on $\Omega$. In (33b) $e_n$ are the elements of the orthonormal base of measurability $Y=\{e_n\}$ and for any $n \leq N(\boldsymbol{w})$ the operator $\hat{q}_n(\boldsymbol{w})$ is a one-dimensional projection on the Hilbert space $H(\boldsymbol{w})$. Every decomposable operator on the Hilbert space $H(\boldsymbol{n},N;Y)$ commutes [20] with the projection-valued measure $\hat{X}_{\boldsymbol{n}}(\cdot)$, hence, in particular, we have

$$[\hat{Q}_n, \hat{X}_{\boldsymbol{n}}(\cdot)] = \hat{0}. \tag{33c}$$

Since the following relations are valid:

$$\hat{q}_n(\boldsymbol{w})\hat{q}_m(\boldsymbol{w}) = \boldsymbol{d}_{nm}\hat{q}_n, \quad \sum_{n=1}^{N(\boldsymbol{w})} \hat{q}_n(\boldsymbol{w}) = \hat{I}_{H(\boldsymbol{w})}, \tag{33d}$$

$\boldsymbol{n}$-a.e. on $\Omega$, the projections $\hat{Q}_n$ are mutually orthogonal and

$$\sum_{n=1}^{l} \hat{Q}_n = \hat{I}_{H(\boldsymbol{n},N;Y)}. \tag{33e}$$

If $\hat{Z}$ is any unitary decomposable operator on $H(\boldsymbol{n},N;Y)$, described by (28g), then the projections

$$\hat{Q}_n^{(Z)} = \hat{Z}\hat{Q}_n\hat{Z}^+, \quad n=1,...,l, \tag{33f}$$

presented by the relation

$$(\hat{Q}_n^{(Z)} g)(\boldsymbol{w}) = ((\hat{Z}\hat{Q}_n\hat{Z}^+)g)(\boldsymbol{w}) = z(\boldsymbol{w})\hat{q}_n\hat{z}^+(\boldsymbol{w})g(\boldsymbol{w}), \tag{33g}$$

$\boldsymbol{n}$-a. e. on $\Omega$, are also mutually orthogonal and summing up to the unity operator: $\sum_{n=1}^{l} \hat{Q}_n^{(Z)} = \hat{I}_{H(\boldsymbol{n},N;Y)}$.

We denote

$$\hat{z}(\boldsymbol{w})\hat{q}_n\hat{z}^+(\boldsymbol{w}) = \hat{q}_n^{(z)} = e_n^{(z)} < e_n^{(z)}, \cdot >_{H(\boldsymbol{w})}. \tag{33h}$$

For any $\boldsymbol{w}\in\Omega$ the set $\{e_n^{(z)}(\boldsymbol{w}), n=1,...,N(\boldsymbol{w})\}$ represents the new orthonormal basis in $H(\boldsymbol{w})$ and the following relation is valid:

$$e_n^{(z)}(\boldsymbol{w}) = \hat{z}(\boldsymbol{w})e_n(\boldsymbol{w}) = \sum_{m=1}^{N(\boldsymbol{w})} \boldsymbol{z}_{nm}^{(z)}(\boldsymbol{w})e_m(\boldsymbol{w}), \tag{33i}$$

where the complex-valued $\boldsymbol{n}$-measurable functions

$$\boldsymbol{z}_{nm}^{(z)}(\boldsymbol{w}) = < e_m(\boldsymbol{w}), \hat{z}(\boldsymbol{w})e_n(\boldsymbol{w}) >_{H(\boldsymbol{w})}, \tag{33j}$$

are the elements of a unitary matrix $\{\boldsymbol{z}_{nm}^{(z)}(\boldsymbol{w})\}$. This matrix may be infinite.

The probability scalar measures (31a), invariant for the class $G_g$, can be now represented in the form:

(34a) $$\boldsymbol{n}_g^{(i)}(E) = tr[\hat{\boldsymbol{S}}_n^{(i)} \hat{X}_n(E)] = (k_g(\boldsymbol{a}_g^{(i)}))^{-1} \int_{\boldsymbol{w} \in E} \sum_{\substack{k=1,\ldots,k_g(\boldsymbol{a}_g^{(i)}), \\ n=1,\ldots,N(\boldsymbol{w})}} |<e_n(\boldsymbol{w}), \boldsymbol{j}_{ik}(\boldsymbol{w}) >_{H(\boldsymbol{w})}|^2 \boldsymbol{n}(d\boldsymbol{w}),$$

(34b) $$\boldsymbol{n}_g(E) = tr[\hat{\boldsymbol{S}}_n \hat{X}_n(E)] = \sum_{i=1}^{N_g} \boldsymbol{a}_g^{(i)} k_g(\boldsymbol{a}_g^{(i)}) \boldsymbol{n}_g^{(i)}(E).$$

The integrand sum in (34a) is invariant under the the transforms (33a,i), that is, we have

(34c) $$\sum_{\substack{k=1,\ldots,k_g(\boldsymbol{a}_g^{(i)}), \\ n=1,\ldots,N(\boldsymbol{w})}} |<e_n(\boldsymbol{w}), \boldsymbol{j}_{ik}(\boldsymbol{w}) >_{H(\boldsymbol{w})}|^2 = \sum_{\substack{k=1,\ldots,k_g(\boldsymbol{a}_g^{(i)}), \\ n=1,\ldots,N(\boldsymbol{w})}} |<e_n^{(z)}(\boldsymbol{w}), \tilde{\boldsymbol{j}}_{ik}(\boldsymbol{w}) >_{H(\boldsymbol{w})}|^2,$$

$\boldsymbol{n}$-a.e. on $\Omega$.

For further consideration we prove some lemmas.

**Lemma 1.** *Let $H(\boldsymbol{n}, N; Y)$ be a direct integral with a finite positive scalar measure $\boldsymbol{n}(\cdot)$. For any unitary operator $\hat{U}$ on $H_S \otimes H(\boldsymbol{n}, N; Y)$ and any unit vector $\boldsymbol{h}_i \in H(\boldsymbol{n}, N; Y)$, there exists a uniquely determined $\boldsymbol{n}$-measurable operator-valued function $\hat{K}_{in}^{(U)}(\cdot)$ on $\Omega$ such that:*

• *For $\forall E \in F_B$*

(35a) $$\int_{\boldsymbol{w} \in E} \hat{K}_{in}^{(U)}(\boldsymbol{w}) \boldsymbol{n}(d\boldsymbol{w})$$

*is a bounded linear operator on $H_S$;*

• *The relation*

(35b) $$((\hat{I} \otimes \hat{Q}_n \hat{X}_n(E)) \hat{U}(g \otimes \boldsymbol{h}_i))(\boldsymbol{w}) = \boldsymbol{c}_E(\boldsymbol{w})(\hat{K}_{in}^{(U)}(\boldsymbol{w}) g \otimes e_n(\boldsymbol{w})), \quad \forall g \in H_S, \quad \forall E \in F_B,$$

*is valid $\boldsymbol{n}$-almost everywhere on $\Omega$, where in (35b) the index n is less or equal to $N(\boldsymbol{w})$;*

• *For $n > N(\boldsymbol{w})$ the operator-valued function $\hat{K}_{in}^{(U)}(\boldsymbol{w}) = \hat{0}$ $\boldsymbol{n}$-almost everywhere on $\Omega$;*

• *The operator-valued function $\hat{K}_{in}^{(U)}(\boldsymbol{w})$ is the Radon-Nikodym derivative of the operator-valued measure*

(35c) $$\mathrm{E}_{|\boldsymbol{h}_i><\boldsymbol{h}_i|}[(\hat{I} \otimes \hat{Q}_n \hat{X}(E)) \hat{U}]$$

*with respect to the finite complex scalar measure $\boldsymbol{m}_{h_i}(d\boldsymbol{w}) = <\boldsymbol{h}_i(\boldsymbol{w}), e_n(\boldsymbol{w}) > \boldsymbol{n}(d\boldsymbol{w})$.*

• *The relation*

(35d) $$(\hat{\mathrm{K}}_{in} g)(\boldsymbol{w}) = \hat{K}_{in}^{(U)}(\boldsymbol{w}) g, \quad \forall g \in H_S,$$

*holding $\boldsymbol{n}$-a. e., defines the bounded linear operator $\hat{\mathrm{K}}_{in} : H_S \to L_2(\Omega, \boldsymbol{n}; H_S)$ with the norm $\|\hat{\mathrm{K}}_{in}\| \leq 1$.*

**Proof.** Let $\{\boldsymbol{f}_k\}$ and $\{\boldsymbol{x}_j\}$ be any complete systems of orthonormal vectors in $H_S$ and $H(\boldsymbol{n}, N; Y)$, respectively. Then the following relation is valid

(36a) $$\hat{U}(g \otimes \boldsymbol{h}_i) = \sum_{j,k} c_{jk}(\boldsymbol{f}_k \otimes \boldsymbol{x}_j) = \sum_j \hat{K}_j(U, \boldsymbol{h}_i) g \otimes \boldsymbol{x}_j, \quad \forall g \in H,$$

where $\hat{K}_j(U, \boldsymbol{h}_i)$ is uniquely determined bounded linear operator on $H_S$ with the norm $\left\|\hat{K}_j(U, \boldsymbol{h}_i)\right\|_{H_S} \leq 1$, such that the relation

(36b) $$<f\otimes \mathbf{x}_j,\hat{U}(g\otimes \mathbf{h}_i)>=<f,\hat{K}_j(\hat{U},\mathbf{h}_i)g>,$$

is valid for any $\forall f,g\in H_S$. In physical notation

(36c) $$\hat{K}_j(U,\mathbf{h}_i)=<\mathbf{x}_j,\hat{U}\mathbf{h}_i>_{H(\mathbf{n},N;Y)},\quad j=1,2,...$$

In the considered case when the measure $\mathbf{n}(\cdot)$ in the direct integral $H(\mathbf{n},N;Y)$ is finite, all elements of the ortonormal base of measurability $Y=\{e_n\}$ belong to $H(\mathbf{n},N;Y)$. From (36a) it follows then that for any vectors $g,f\in H_S$, for any $E\in F_B$ and any index $n=1,...,l$, where $l$ is equal to $\mathbf{n}\text{-}\sup\{N(\mathbf{w}),\mathbf{w}\in\Omega\}$, we have

(36d) $$<f\otimes e_n,(\hat{I}\otimes \hat{X}_\mathbf{n}(E))\hat{U}(g\otimes \mathbf{h}_i)>=\int_{\mathbf{w}\in E}<f,\{\sum_j \hat{K}_j(U,\mathbf{h}_i)<e_n(\mathbf{w}),\mathbf{x}_j(\mathbf{w})>_{H(\mathbf{w})}\}g>_{H_S}\mathbf{n}(d\mathbf{w}),$$

where for the left hand side in (36d) the following bound is valid:

(36e) $$|<f\otimes e_n,(\hat{I}\otimes \hat{X}_\mathbf{n}(E))\hat{U}(g\otimes \mathbf{h}_i)>|\le \sqrt{\mathbf{n}(E)}\,\|f\|\,\|g\|,\quad \forall f,g\in H_S,\ \forall E\in F_B.$$

Due to (36d) for any $E\in F_B$ the sequence

(36f) $$\int_{\mathbf{w}\in E}<f,\{\sum_{j=1}^{m}\hat{K}_j(U,\mathbf{h}_i)<e_n(\mathbf{w}),\mathbf{x}_j(\mathbf{w})>_{H(\mathbf{w})}\}g>_{H_S}\mathbf{n}(d\mathbf{w})=$$
$$=<f,\{\int_{\mathbf{w}\in E}\sum_{j=1}^{m}\hat{K}_j(U,\mathbf{h}_i)<e_n(\mathbf{w}),\mathbf{x}_j(\mathbf{w})>_{H(\mathbf{w})}\mathbf{n}(d\mathbf{w})\}g>_{H_S}$$

converges as $m\to\infty$ for any $g,f\in H_S$. Consequently, there exists [23] the bounded linear operator on $H_S$, which we denote

(36g) $$\int_{\mathbf{w}\in E}\hat{K}_{in}^{(U)}(\mathbf{w})\mathbf{n}(d\mathbf{w}),$$

such that for any $E\in F_B$

(36h) $$\int_{\mathbf{w}\in E}\sum_{j=1}^{m}\hat{K}_j(U,\mathbf{h}_i)<e_n(\mathbf{w}),\mathbf{x}_j(\mathbf{w})>_{H(\mathbf{w})}\mathbf{n}(d\mathbf{w})\xrightarrow{W}\int_{\mathbf{w}\in E}\hat{K}_{in}^{(U)}(\mathbf{w})\mathbf{n}(d\mathbf{w}).$$

as $m\to\infty$.

For the operator-valued densities in (36h) we have the relation

(36i) $$<f,\{\sum_{j=1}^{m}\hat{K}_j(U,\mathbf{h}_i)<e_n(\mathbf{w}),\mathbf{x}_j(\mathbf{w})>_{H(\mathbf{w})}\}g>_{H_S}\xrightarrow{W}<f,\hat{K}_{in}^{(U)}(\mathbf{w})g>_{H_S}$$

as $m\to\infty$, which is valid for any $f,g\in H_S$.

Thus, for any $g,f\in H_S$ we can rewrite (36d) in the form

(37a) $$<f\otimes e_n,(\hat{I}\otimes \hat{X}_\mathbf{n}(E))\hat{U}(g\otimes \mathbf{h}_i)>=\int_{\mathbf{w}\in E}<f,\hat{K}_{in}^{(U)}(\mathbf{w})g>_{H_S}\mathbf{n}(d\mathbf{w}),$$

with $\hat{K}_{in}^{(U)}(\cdot)$ being a $\mathbf{n}$-measurable function on $\Omega$ with values being linear operators on $H_S$ defined for any $g\in H_S$ $\mathbf{n}$-almost everywhere on $\Omega$ and such that (36g) is a bounded linear operator on $H_S$ for any $E\in F_B$.

From (36i) it follows that $\hat{K}_{in}^{(U)}(\mathbf{w})=\hat{0}$ for $n>N(\mathbf{w})$, $\mathbf{n}$-a.e. on $\Omega$.

The bounded operator adjoint to the bounded operator (36g) is given by

(37b) $$\int_{\mathbf{w}\in E}(\hat{K}_{in}^{(U)}(\mathbf{w}))^+\mathbf{n}(d\mathbf{w})$$

with the operator-valued density $(\hat{K}_{in}^{(U)}(\mathbf{w}))^+$, satisfying the relation

(37c) $\quad < f, \{\sum_{j=1}^{m}(\hat{K}_j(U, \mathbf{h}_i))^+ < \mathbf{x}_j(\mathbf{w}), e_n(\mathbf{w}) >_{H(\mathbf{w})}\} g >_{H_S} \xrightarrow{W} < f, (\hat{K}_{in}^{(U)}(\mathbf{w}))^+ g >_{H_S} \qquad m \to \infty$

for any $f, g \in H_S$.

Due to (28d) and (33b) and (36a,i) we also have that for any $g \in H_S$, $\forall n \leq N(\mathbf{w})$, $\forall E \in F_B$ the relation

(37d) $\quad ((\hat{I} \otimes \hat{Q}_n \hat{X}_{\mathbf{n}}(E))\hat{U}(g \otimes \mathbf{h}_i))(\mathbf{w}) = \mathbf{c}_E(\mathbf{w})(\hat{K}_{in}^{(U)}(\mathbf{w})g \otimes e_n(\mathbf{w}))$,

is valid $\mathbf{n}$-a.e. on $\Omega$.

Substituting (37d) into (35c), we get the following relation:

(37e) $\quad E_{|h_i > < h_i|}[(\hat{I} \otimes \hat{Q}_n \hat{X}_{\mathbf{n}}(E))\hat{U}] = \int_{\mathbf{w} \in E} \hat{K}_{in}^{(U)}(\mathbf{w}) < \mathbf{h}_i(\mathbf{w}), e_n(\mathbf{w}) >_{H(\mathbf{w})} \mathbf{n}(d\mathbf{w})$

and, consequently, the operator-valued function $\hat{K}_{in}^{(U)}(\mathbf{w})$ is the Radon-Nikodym derivative of the operator-valued measure (35c) with respect to the finite complex scalar measure

(37f) $\quad \mathbf{m}_{h_i}(d\mathbf{w}) = < \mathbf{h}_i(\mathbf{w}), e_n(\mathbf{w}) > \mathbf{n}(d\mathbf{w})$.

The last statement of the lemma is based on the fact that from (37d) it follows that in case $E = \Omega$ we have

(37g) $\quad \int_\Omega \sum_{n=1}^{N(\mathbf{w})} < \hat{K}_{in}^{(U)}(\mathbf{w})\mathbf{y}, \hat{K}_{in}^{(U)}(\mathbf{w})\mathbf{y} >_{H_S} \mathbf{n}(d\mathbf{w}) = \|\mathbf{y}\|^2, \qquad \forall \mathbf{y} \in H_S$.

It is easy to prove also the following lemma.

**Lemma 2.** *Let $\mathbf{h}_i, \mathbf{h}_j \in H(\mathbf{n}, N)$ be some orthogonal unit vectors, then the $\mathbf{n}$-measurable operator-valued functions $\hat{K}_{in}^{(U)}(\mathbf{w})$ and $\hat{K}_{jn}^{(U)}(\mathbf{w})$, given by (35) for unit vectors $\mathbf{h}_i$ and $\mathbf{h}_j$, respectively, satisfy the following orthonormality relation*

(38) $\quad \int_\Omega \sum_{n=1}^{N(\mathbf{w})} (\hat{K}_{jn}^{(U)}(\mathbf{w}))^+ \hat{K}_{in}^{(U)}(\mathbf{w}) \mathbf{n}(d\mathbf{w}) = \mathbf{d}_{ji} \hat{I}$.

Consider now the expressions for the operator-valued measures (32a), which are (up to phase equivalence) invariants for the class $G_g$.

**Definition.** *For any indexes $i = 1,...,N_g$, $k = 1,...,k_g(\mathbf{a}_g^{(i)})$ and $n = 1,...,l$, where the positive integer $l$ is equal to $\mathbf{n}$-$\sup\{N(\mathbf{w}), \mathbf{w} \in \Omega\}$, define for any $\mathbf{y} \in H_S$ the $\mathbf{n}$-measurable operator-valued function $\hat{V}_{in}^{(k)}(\mathbf{w})$ by the relation*

(39a) $\quad ((\hat{I} \otimes \hat{Q}_n \hat{X}_{\mathbf{n}}(E))\hat{U}_{\mathbf{n}}(\mathbf{y} \otimes \mathbf{j}_{ik}))(\mathbf{w}) = \mathbf{c}_E(\mathbf{w})(\hat{V}_{in}^{(k)}(\mathbf{w})\mathbf{y} \otimes e_n(\mathbf{w}))$,

$\mathbf{n}$-*a. e. on* $\Omega$. *Here* $\forall n \leq N(\mathbf{w})$.

The correctness of this definition follows from lemma 1.

It follows from (39a) that under the unitary transforms (33a,i) the operators $\hat{V}_{in}^{(k)}(\mathbf{w})$ are transformed as

(39b) $\quad \hat{V}_{in}^{(k)}(\mathbf{w}) = \sum_{\substack{p=1,...,k_g(\mathbf{a}_g^{(i)}) \\ m=1,...,N(\mathbf{w})}} \mathbf{z}_{mn}^{(z)}(\dot{u}) \mathbf{J}_{kp}^{(i)} \hat{\tilde{V}}_{im}^{(p)}(\dot{u})$.

From lemma 2 we have the following orthonormality relation for operators $\hat{V}_{in}^{(k)}(\boldsymbol{w})$ :

(39c) $$\int_\Omega \sum_{n=1}^{N(\boldsymbol{w})} (\hat{V}_{jn}^{(p)}(\boldsymbol{w}))^+ \hat{V}_{in}^{(k)}(\boldsymbol{w})\boldsymbol{n}(d\boldsymbol{w}) = \boldsymbol{d}_{ji}\boldsymbol{d}_{pk}\hat{I}.$$

Considering (35b), (31d) and (39a), we get the following integral representations for the operator-valued measures (32a,b)

(40a) $$\hat{\Theta}_{\tilde{a}}^{(i)}(E) = \mathrm{E}_{\hat{s}_{\boldsymbol{n}}^{(i)}}[(\hat{I}\otimes \hat{X}_{\boldsymbol{n}}(E))\hat{U}_{\boldsymbol{n}}] = (k_g(\boldsymbol{a}_g^{(i)}))^{-1} \int_{\boldsymbol{w}\in E} \sum_{\substack{n=1,...,N(\boldsymbol{w})\\ k=1,...,k_g(\boldsymbol{a}_g^{(i)})}} \hat{V}_{in}^{(k)}(\boldsymbol{w}) <\boldsymbol{j}_{ik}(\boldsymbol{w}),e_n(\boldsymbol{w})>_{H(\boldsymbol{w})} \boldsymbol{n}(d\boldsymbol{w}),$$

(40b) $$\hat{\Theta}_{\tilde{a}}(E) = \mathrm{E}_{\hat{s}_{\boldsymbol{n}}}[(\hat{I}\otimes \hat{X}_{\boldsymbol{n}}(E))\hat{U}_{\boldsymbol{n}}] = \sum_{i=1}^{N_g} \boldsymbol{a}_g^{(i)} \int_{\boldsymbol{w}\in E} \sum_{\substack{n=1,...,N(\boldsymbol{w})\\ k=1,...,k_g(\boldsymbol{a}_g^{(i)})}} \hat{V}_{in}^{(k)}(\boldsymbol{w}) <\boldsymbol{j}_{ik}(\boldsymbol{w}),e_n(\boldsymbol{w})>_{H(\boldsymbol{w})} \boldsymbol{n}(d\boldsymbol{w}).$$

The integral representations for the "$i$" instruments (30c), which are invariants for the classes $G_g, [G_g]$, have the form

(41a) $$\hat{T}_i(E)[\hat{A}] = (k_g(\boldsymbol{a}_g^{(i)}))^{-1} \int_{\boldsymbol{w}\in E} \sum_{\substack{n=1,...,N(\boldsymbol{w})\\ k=1,...,k_g(\boldsymbol{a}_g^{(i)})}} (\hat{V}_{in}^{(k)}(\boldsymbol{w}))^+ \hat{A}\, \hat{V}_{in}^{(k)}(\boldsymbol{w})\boldsymbol{n}(d\boldsymbol{w}),$$

while for the whole instrument we have

(41b) $$\hat{T}(E)[\hat{A}] = \sum_{i=1,...,N_g} \boldsymbol{a}_g^{(i)} \int_{\boldsymbol{w}\in E} \sum_{\substack{n=1,...,N(\boldsymbol{w})\\ k=1,...,k_g(\boldsymbol{a}_g^{(i)})}} (\hat{V}_{in}^{(k)}(\boldsymbol{w}))^+ \hat{A}\hat{V}_{in}^{(k)}(\boldsymbol{w})\boldsymbol{n}(d\boldsymbol{w}).$$

The integrand sums in (40a,b) and (41a,b) are invariant under the the transforms (33a,i), that is, $\boldsymbol{n}$-a.e. on $\Omega$

(42a) $$\sum_{\substack{n=1,...,N(\boldsymbol{w})\\ k=1,...,k_g(\boldsymbol{a}_g^{(i)})}} \hat{V}_{in}^{(k)}(\boldsymbol{w}) <\boldsymbol{j}_{ik}(\boldsymbol{w}),e_n(\boldsymbol{w})>_{H(\boldsymbol{w})} = \sum_{\substack{n=1,...,N(\boldsymbol{w})\\ k=1,...,k_g(\boldsymbol{a}_g^{(i)})}} \hat{\tilde{V}}_{in}^{(k)}(\boldsymbol{w}) <\tilde{\boldsymbol{j}}_{ik}(\boldsymbol{w}),e_n^{(z)}(\boldsymbol{w})>_{H(\boldsymbol{w})}$$

and

(42b) $$\sum_{\substack{n=1,...,N(\boldsymbol{w})\\ k=1,...,k_g(\boldsymbol{a}_g^{(i)})}} (\hat{V}_{in}^{(k)}(\boldsymbol{w}))^+ \hat{A}\hat{V}_{in}^{(k)}(\boldsymbol{w}) = \sum_{\substack{n=1,...,N(\boldsymbol{w})\\ k=1,...,k_g(\boldsymbol{a}_g^{(i)})}} (\hat{\tilde{V}}_{in}^{(k)}(\boldsymbol{w}))^+ \hat{A}\hat{\tilde{V}}_{in}^{(k)}(\boldsymbol{w}).$$

If we denote

(43a) $$q_{in}^{(k)}(\boldsymbol{w}) = <e_n(\boldsymbol{w}),\boldsymbol{j}_{ik}(\boldsymbol{w})>_{H(\boldsymbol{w})},$$

then in (34a) the densities $\boldsymbol{p}_i(\boldsymbol{w})$ of the probability scalar measures $\boldsymbol{n}_g^{(i)}(\cdot)$, $i=1,...,N_g$ with respect to the finite positive scalar measure $\boldsymbol{n}(\cdot)$ can be represented as

(43b) $$\boldsymbol{p}_i(\boldsymbol{w}) = (k_g(\boldsymbol{a}_g^{(i)}))^{-1} \sum_{\substack{n=1,...,N(\boldsymbol{w})\\ k=1,...,k_g(\boldsymbol{a}_g^{(i)})}} |q_{in}^{(k)}(\boldsymbol{w})|^2$$

and, consequently,

(43c) $$\boldsymbol{n}_g^{(i)}(d\boldsymbol{w}) = (k_g(\boldsymbol{a}_g^{(i)}))^{-1} \sum_{\substack{n=1,...,N(\boldsymbol{w})\\ k=1,...,k_g(\boldsymbol{a}_g^{(i)})}} |q_{in}^{(k)}(\boldsymbol{w})|^2 \boldsymbol{n}(d\boldsymbol{w}).$$

The operator-valued measures (40a) can be rewritten as

(43d) $$\hat{\Theta}_{\tilde{a}}^{(i)}(E) = \mathrm{E}_{\hat{s}_{\boldsymbol{n}}^{(i)}}[(\hat{I}\otimes \hat{X}_{\boldsymbol{n}}(E))\hat{U}_{\boldsymbol{n}}] = (k_g(\boldsymbol{a}_g^{(i)}))^{-1} \int_{\boldsymbol{w}\in E} \sum_{\substack{n=1,...,N(\boldsymbol{w})\\ k=1,...,k_g(\boldsymbol{a}_g^{(i)})}} \hat{V}_{in}^{(k)}(\boldsymbol{w})(q_{in}^{(k)}(\boldsymbol{w}))^* \boldsymbol{n}(d\boldsymbol{w})$$

The following orthonormality relation is valid for the scalar products, introduced by (43a):

(43e)
$$\int_\Omega \sum_{n=1,...,N(\boldsymbol{w})} (q_{jn}^{(k)}(\boldsymbol{w}))^* q_{in}^{(p)}(\boldsymbol{w}) \boldsymbol{n}(d\boldsymbol{w}) = \boldsymbol{d}_{ji} \boldsymbol{d}_{kp}.$$

We also have

(43f)
$$\boldsymbol{n}_g^{(i)} \prec \boldsymbol{n}_g \prec \boldsymbol{n},$$

where $[\boldsymbol{n}(\cdot)] = [\hat{P}_R(\cdot)]$. Recall that in (43f) only the probability scalar measures $\boldsymbol{n}_g^{(i)}(\cdot)$, $i = 1,..., N_g$ and $\boldsymbol{n}_g(\cdot)$ are invariants of the class $G_g$.

Let introduce the double index $(i,k) \to i$ and repeat in the sum in (41b) the number $\boldsymbol{a}_i$ as many times as its multiplicity $k(\boldsymbol{a}_i)$ is. Then we can present our result in the form of the following theorem. In this theorem no reference is made to invariance properties, the invariance aspects are taken up again in the next sections.

**Theorem 1 (The most general form of an integral representation of an instrument).**
*Let $(\Omega, F_B)$ be a standard Borel space. For any instrument $\hat{T}(\cdot)[\cdot]: F_B \times L(H_S) \to L(H_S)$ there exist:*
- *a finite positive scalar measure $\boldsymbol{n}(\cdot)$ on $(\Omega, F_B)$;*
- *a family $\{\hat{V}_{in}(\boldsymbol{w}), \boldsymbol{w} \in \Omega;\ i = 1,..., N_0; n = 1,..., N(\boldsymbol{w})\}$ of $\boldsymbol{n}$-measurable operator-valued functions on $\Omega$, such that for any $\forall E \in F_B$ and any indexes $i, n$*

(44a)
$$\int_{\boldsymbol{w} \in E} \hat{V}_{in}(\boldsymbol{w}) \boldsymbol{n}(d\boldsymbol{w})$$

*is a bounded linear operator on $H_S$ and the following orthonormality relation is valid:*

(44b)
$$\int_\Omega \sum_{n=1,...,N(\boldsymbol{w})} \hat{V}_{jn}^+(\boldsymbol{w}) \hat{V}_{in}(\boldsymbol{w}) \boldsymbol{n}(d\boldsymbol{w}) = \boldsymbol{d}_{ji} \hat{I};$$

- *a sequence of positive numbers $\{\boldsymbol{a}_1, \boldsymbol{a}_2,...\}$, satisfying*

(44c)
$$\sum_{i=1}^{N_0} \boldsymbol{a}_i = 1, \quad N_0 \leq \infty;$$

*such that the instrument can be presented as*

(44d)
$$\hat{T}(E)[\hat{A}] = \sum_{i=1,...,N_0} \boldsymbol{a}_i \int_{\boldsymbol{w} \in E} \sum_{n=1}^{N(\boldsymbol{w})} \hat{V}_{in}^+(\boldsymbol{w}) \hat{A} \hat{V}_{in}(\boldsymbol{w}) \boldsymbol{n}(d\boldsymbol{w})$$

*on all of $H_S$ for $\forall E \in F_B, \forall \hat{A} \in L(H_S)$.*
*Furthermore, for any indexes $i, n$ the relation*

(44e)
$$(\hat{W}_{in} \boldsymbol{y})(\boldsymbol{w}) = \hat{V}_{in}(\boldsymbol{w}) \boldsymbol{y}, \qquad \forall \boldsymbol{y} \in H_S,$$

*holding $\boldsymbol{n}$-a.e. on $\Omega$, defines the bounded linear operator $\hat{W}_{in}: H_S \to L_2(\Omega, \boldsymbol{n}; H_S)$ with the norm $\|\hat{W}_{in}\| \leq 1$.*

In the general case the form of the integral representation of an instrument (44a)-(44d) differs from that given by (20) [cf.10,12], where the orthogonality relation (44b) is not present. The double index in (44a)-(44d) (in comparison with the single index in (20)) can not be presented as a single one since these indexes enter the orthonormality relation (44b) in different manner. In contradistinction to the arguments presented in [12], we prove (see lemma 1) that the representation (44d) through operator-valued functions $\hat{V}_{in}(\boldsymbol{w})$ is valid on all of $H_S$.

In the case of discrete character of the measure $\boldsymbol{n}(\cdot)$ the representation (44d), due to the orthogonality relation (44b), differs also from formulae, of a somewhat similar form, available in the physical literature [18,19]

*3.3. Stochastic realizations of an instrument*

For the definite index $i$ the operators $\hat{V}_{in}^{(k)}(\boldsymbol{w})$ in (41b) are defined with respect to the concrete choice of the basis $\{\boldsymbol{j}_{ik}, k=1,...,k_g(\boldsymbol{a}_g^{(i)})\}$ in the subspace $\hat{p}_i H(\boldsymbol{n}, N; Y)$ and the concrete decomposition (33e) of the unity operator on $H(\boldsymbol{n}, N; Y)$. If in the statistical realization $\boldsymbol{g}$, given by (24b) and unitarily equivalent to the statistical realization (28e) on $H(\boldsymbol{n}, N; Y)$, the density operator $\hat{\boldsymbol{s}}_R$ has eigenvalues $\boldsymbol{a}_g^{(i)}$ with multiplicity $k_g(\boldsymbol{a}_g^{(i)}) \neq 1$, as well as if in the statistical realization $\boldsymbol{g}$ the projection-valued measure $\hat{P}_R(\cdot)$ on $(\Omega, F_B)$ is not simple (that is, for some $\boldsymbol{w} \in \Omega$, $N_g(\boldsymbol{w}) \neq 1$), then even for the definite statistical realization we have a plenitude of integral representations (44d) of the corresponding instrument.

Moreover, the operators $\hat{V}_{in}^{(k)}(\boldsymbol{w})$ are defined with respect to the concrete finite positive scalar measure $\boldsymbol{n}(\cdot)$ from the equivalence class $[\boldsymbol{n}(\cdot)] = [\hat{X}(\cdot)] = [\hat{P}_R(\cdot)]$. Although $[\boldsymbol{n}(\cdot)] = [\hat{X}(\cdot)] = [\hat{P}_R(\cdot)]$ is an invariant of the classes $G_g, [G_g]$, the measure $\boldsymbol{n}(\cdot)$ itself is not an invariant of these classes and can be chosen in many ways.

Thus, in (41b) the positive scalar measure $\boldsymbol{n}(\cdot)$ and operators $\hat{V}_{in}^{(k)}(\boldsymbol{w})$, defined with respect to this measure, are not invariants of the classes $G_g, [G_g]$ of statistical realizations.

However, there is a definite correspondence between the classes $G_g, [G_g]$ of statistical realizations and integral representations of the instrument, corresponding to these classes. In this section we analyse this correspondence.

Introduce the following definition.

**Definition.** *Consider a triple $\boldsymbol{l} = \{\boldsymbol{b}_l, \Lambda_l, V_l\}$, consisting of:*

• *a family $\boldsymbol{b}_l$ of positive coefficients $\boldsymbol{b}_l^{(i)} > 0$, summing up to identity, where every coefficient may be repeated $k_l(\boldsymbol{b}_l^{(i)})$ times:*

(45a) $$\boldsymbol{b}_l = \{\{\boldsymbol{b}_l^{(i)}, k_l(\boldsymbol{b}_l^{(i)})\} / \boldsymbol{b}_l^{(i)} > 0, i = 1,..., \tilde{N}_l; \sum_i \boldsymbol{b}_l^{(i)} k_l(\boldsymbol{b}_l^{(i)}) = 1\};$$

• *a family $\Lambda_l$ of complex scalar measures, absolutely continuous with respect to a finite positive scalar measure $\tilde{\boldsymbol{n}}(\cdot)$ and satisfying the orthonormality relation:*

(45b) $$\Lambda_l = \{\tilde{q}_{in}^{(k)}(\boldsymbol{w}) \tilde{\boldsymbol{n}}(d\boldsymbol{w}) | \boldsymbol{w} \in \Omega; i = 1,..,\tilde{N}_l; k = 1,..,k_l(\boldsymbol{b}_l^{(i)}); n = 1,..,\tilde{N}(\boldsymbol{w});$$
$$\int_\Omega \sum_{n=1}^{\tilde{N}(\boldsymbol{w})} (\tilde{q}_{jn}^{(k)}(\boldsymbol{w}))^* \tilde{q}_{in}^{(p)}(\boldsymbol{w}) \tilde{\boldsymbol{n}}(d\boldsymbol{w}) = \boldsymbol{d}_{kp} \boldsymbol{d}_{ji}\},$$

*where in (45a,b) the positive integers $\tilde{N}_l, \tilde{N}(\boldsymbol{w})$ may be infinite;*

• *a family $V_l$ of $\tilde{\boldsymbol{n}}$-measurable operator-valued functions $\hat{W}_{in}^{(k)}(\cdot)$ on $\Omega$, satisfying the orthonormality relation,*

(45c) $$V_l = \{\hat{W}_{in}^{(k)}(\boldsymbol{w}) | \boldsymbol{w} \in \Omega, i = 1,...,\tilde{N}_l; k = 1,...,k_l(\boldsymbol{b}_l^{(i)}); n = 1,...\tilde{N}(\boldsymbol{w});$$
$$\int_\Omega \sum_{n=1}^{\tilde{N}(\boldsymbol{w})} (\hat{W}_{jn}^{(k)}(\boldsymbol{w}))^+ \hat{W}_{in}^{(p)}(\boldsymbol{w}) \tilde{\boldsymbol{n}}(d\boldsymbol{w}) = \boldsymbol{d}_{kp} \boldsymbol{d}_{ji} \hat{I}\}$$

*and such that for all indexes $i, n, k$ and any $E \in F_B$*

(45d) $$\int_{\boldsymbol{w} \in E} \hat{W}_{in}^{(k)}(\boldsymbol{w}) \tilde{\boldsymbol{n}}(d\boldsymbol{w}), \quad \int_{\boldsymbol{w} \in E} \sum_{\substack{n=1,...,\tilde{N}(\boldsymbol{w}),\\k=1,...,k_l(\boldsymbol{b}_l^{(i)})}} \hat{W}_{in}^{(k)}(\boldsymbol{w})(q_{in}^{(k)}(\boldsymbol{w}))^* \tilde{\boldsymbol{n}}(d\boldsymbol{w})$$

*are bounded linear operators on $H_S$;*

*We shall call $\mathbf{1} = \{\mathbf{b}_1, \Lambda_1, V_1\}$ a stochastic realization of an instrument, if the instrument can be represented in the integral form:*

$$\hat{T}(E)[\hat{A}] = \sum_{i,k} \mathbf{b}_1^{(i)} \int_{\mathbf{w} \in E} \sum_{n=1}^{\tilde{N}(\mathbf{w})} (\hat{W}_{in}^{(k)}(\mathbf{w}))^+ \hat{A} \hat{W}_{in}^{(k)}(\mathbf{w}) \tilde{\mathbf{n}}(d\mathbf{w}). \tag{45e}$$

*on all of $H_S$. We shall say that we have different stochastic realizations of the same instrument if the triples $\{\mathbf{b}, \Lambda, V\}$ are different.*

The following statement follows from the consideration, presented in section 3.2 and, in particular, from the formulae (41) - (43).

**Proposition 1.** *Let $(\Omega, F_B)$ be a standard Borel space. For any instrument $\hat{T}(\cdot)[\cdot]: F_B \times L(H_S) \to L(H_S)$ there exists a stochastic realization.*

Let $\mathbf{1}$ be a stochastic realization of an instrument. Then, for example, any triple $\mathbf{1}'$, the elements of which are connected with the elements of $\mathbf{1}$ by the following transformation $\tilde{\mathbf{n}}$ - a.e:

$$\mathbf{b}_1 = \mathbf{b}_{1'}; \quad \tilde{\mathbf{n}} \sim \tilde{\mathbf{n}}'; \quad \tilde{N}(\mathbf{w}) = \tilde{N}'(\mathbf{w}); \tag{46a}$$

$$\tilde{q}'^{(k)}_{in}(\mathbf{w}) = \sum_{\substack{m=1,\ldots,\tilde{N}(\mathbf{w}) \\ p=1,\ldots,k_1(\mathbf{b}_1^{(i)})}} a_{nm}^{(1)}(\mathbf{w}) b_{kp}^{(1)}(\mathbf{w}) \tilde{q}_{im}^{(p)}(\mathbf{w}) \sqrt{d\tilde{\mathbf{n}}/d\tilde{\mathbf{n}}'}; \tag{46b}$$

$$\hat{W}'^{(k)}_{in}(\mathbf{w}) = \sum_{\substack{m=1,\ldots,\tilde{N}(\mathbf{w}), \\ p=1,\ldots,k_1(\mathbf{b}_1^{(i)})}} a_{nm}^{(2)}(\mathbf{w}) b_{kp}^{(2)}(\mathbf{w}) \hat{W}_{im}^{(p)}(\mathbf{w}) \sqrt{d\tilde{\mathbf{n}}/d\tilde{\mathbf{n}}'}; \tag{46c}$$

where $\{a_{nm}^{(1,2)}(\mathbf{w})\}$ and $\{b_{kp}^{(1,2)}(\mathbf{w})\}$ are any unitary matrices of complex-valued $\tilde{\mathbf{n}}$-measurable functions, is also a stochastic realization of the same instrument.

Introduce now the equivalence relation on the space of all possible stochastic realizations of the given instrument.

We shall say that two stochastic realizations $\mathbf{1}$ and $\mathbf{1}'$ are equivalent if there exist a real number $\mathbf{x}$, a unitary matrix $\{\mathbf{z}_{mn}(\mathbf{w})\}$ of complex-valued measurable functions and unitary matrices $\{\mathbf{J}_{kp}^{(i)}\}$, $i = 1,\ldots,\tilde{N}_1$ of complex numbers such that $\tilde{\mathbf{n}}$-a.e.:

$$\mathbf{b}_1 = \mathbf{b}_{1'}; \quad \tilde{\mathbf{n}} \sim \tilde{\mathbf{n}}'; \quad \tilde{N}(\mathbf{w}) = \tilde{N}'(\mathbf{w}); \tag{47a}$$

$$\tilde{q}_{in}^{(k)}(\mathbf{w}) = \sum_{\substack{m=1,\ldots,\tilde{N}(\mathbf{w}), \\ k=1,\ldots,k_1(\mathbf{b}_1^{(i)})}} \mathbf{z}_{nm}(\mathbf{w}) \mathbf{J}_{kp}^{(i)} \tilde{q}_{im}'^{(p)}(\mathbf{w}) \sqrt{d\tilde{\mathbf{n}}'/d\tilde{\mathbf{n}}}; \tag{47b}$$

$$\hat{W}_{in}^{(k)}(\mathbf{w}) = e^{i\mathbf{x}} \sum_{\substack{m=1,\ldots,\tilde{N}(\mathbf{w}), \\ k=1,\ldots,k_1(\mathbf{b}_1^{(i)})}} \mathbf{z}_{nm}(\mathbf{w}) \mathbf{J}_{kp}^{(i)} \hat{W}_{im}'^{(p)}(\mathbf{w}) \sqrt{d\tilde{\mathbf{n}}'/d\tilde{\mathbf{n}}} \tag{47c}$$

Denote the equivalent class by $S_1$ if it contains the stochastic realization $\mathbf{1}$. Stochastic realizations from the class $S_1$ have the following invariants.

The type $[\tilde{\mathbf{n}}]$, the dimension function $\{N(\mathbf{w}), \mathbf{w} \in \Omega\}$, the family $\mathbf{b}_1$ of positive coefficients, the probability scalar measures

$$\tilde{\mathbf{n}}_1^{(i)}(d\mathbf{w}) = (k_g(\mathbf{b}_1^{(i)}))^{-1} \sum_{\substack{n=1,\ldots,\tilde{N}(\mathbf{w}), \\ k=1,\ldots,k_1(\mathbf{b}_1^{(i)})}} |\tilde{q}_{in}^{(k)}(\mathbf{w})|^2 \tilde{\mathbf{n}}(d\mathbf{w}), \tag{48a}$$

(48b)
$$\tilde{\boldsymbol{n}}_l(d\boldsymbol{w}) = \sum_{i=1,...,\tilde{N}_l} \boldsymbol{b}_l^{(i)} k_l(\boldsymbol{b}_l^{(i)}) \tilde{\boldsymbol{n}}_l^{(i)}(d\boldsymbol{w})$$

and the operator-valued measures

(48c)
$$\hat{\tilde{\Theta}}_l^{(i)}(E) = (k_g(\boldsymbol{b}_l^{(i)}))^{-1} \int_{\boldsymbol{w} \in E} \sum_{\substack{n=1,...,\tilde{N}(\boldsymbol{w}) \\ k=1,...,k_l(\boldsymbol{b}_l^{(i)})}} \hat{W}_{in}^{(k)}(\boldsymbol{w})(\tilde{q}_{in}^{(k)}(\boldsymbol{w}))^* \tilde{\boldsymbol{n}}(d\boldsymbol{w}),$$

(48d)
$$\hat{\tilde{\Theta}}_l(E) = \sum_{i=1,...,\tilde{N}_l} \hat{a}_l^{(i)} k_l(\boldsymbol{b}_l^{(i)}) \hat{\Theta}_l^{(i)}(E),$$

which are invariants of $S_l$ up to phase equivalence.

Thus, we derived the following set of invariants of the equivalence class $S_l$

(49)
$$[\tilde{\boldsymbol{n}}], \{\tilde{N}_l(\boldsymbol{w}), \boldsymbol{w} \in \Omega\}, \boldsymbol{b}_l, \{\tilde{\boldsymbol{n}}_l^{(i)}(\cdot)\}, \tilde{\boldsymbol{n}}_l(\cdot), \{\hat{\tilde{\Theta}}_l^{(i)}(\cdot)\}, \hat{\tilde{\Theta}}_l(\cdot).$$

Introduce also the equivalence relation on the space $S = \{S_l, l \in \Delta\}$ of equivalence classes $S_l$ of stochastic realizations.

We shall say that two equivalence classes $S_l$ and $S_{l'}$ are equivalent if for any stochastic realization $l$ from the class $S_l$ and any stochastic realization $l'$ from the class $S_{l'}$ there exist a real number $\boldsymbol{x}$, a unitary matrix $\{\boldsymbol{z}_{mn}(\boldsymbol{w})\}$ of complex-valued measurable functions and a unitary matrix $\{\boldsymbol{J}_{kp}^{(i)}\}$ of complex numbers such that $\tilde{\boldsymbol{n}}$-a.e.:

(50a)
$$\tilde{N}_l = \tilde{N}_{l'}, \quad k_l(\boldsymbol{b}_l^{(i)}) = k_{l'}(\boldsymbol{b}_l^{(i)}); \forall i = 1,...,\tilde{N}_l;$$
$$\tilde{\boldsymbol{n}} \sim \tilde{\boldsymbol{n}}'; \quad \tilde{N}(\boldsymbol{w}) = \tilde{N}'(\boldsymbol{w});$$

(50b)
$$\hat{W}_{in}^{(k)}(\boldsymbol{w}) = e^{i\boldsymbol{x}} \sum_{\substack{m=1,...,\tilde{N}(\boldsymbol{w}) \\ k=1,...,k_l(\boldsymbol{b}_l^{(i)})}} \boldsymbol{z}_{nm}(\boldsymbol{w}) \boldsymbol{J}_{kp}^{(i)} \hat{W}_{im}^{\prime(p)}(\boldsymbol{w}) \sqrt{d\tilde{\boldsymbol{n}}'/d\tilde{\boldsymbol{n}}};$$

(50c)
$$\tilde{q}_{in}^{(k)}(\boldsymbol{w}) = \sum_{\substack{m=1,...,\tilde{N}(\boldsymbol{w}) \\ k=1,...,k_l(\boldsymbol{b}_l^{(i)})}} \boldsymbol{z}_{nm}(\boldsymbol{w}) \boldsymbol{J}_{kp}^{(i)} \tilde{q}_{im}^{\prime(p)}(\boldsymbol{w}) \sqrt{d\tilde{\boldsymbol{n}}'/d\tilde{\boldsymbol{n}}}.$$

Denote the equivalent class by $[S_l]$ if it contains the class $S_l$.

The class $[S_l]$ has the following set of invariants

(51)
$$[\tilde{\boldsymbol{n}}], \{\tilde{N}_l(\boldsymbol{w}), \boldsymbol{w} \in \Omega\}, \tilde{N}_l, \{k_l(\boldsymbol{b}_l^{(i)})\}\{\tilde{\boldsymbol{n}}_l^{(i)}(\cdot)\}, \{\hat{\tilde{\Theta}}_l^{(i)}(\cdot)\}.$$

Consider now the correspondence between the different classes of stochastic and statistical realizations of the same instrument.

**Theorem 2.** *There is a one-to-one correspondence between the space $S = \{S_l, l \in \Delta\}$ of equivalence classes of stochastic realizations and the space $G = \{G_g, g \in \Gamma\}$ of classes of unitarily equivalent statistical realizations of the instrument. The element of S corresponds to the element of G if and only if they have the same sets of invariants, given by (49) and (32c), respectively.*
*Every statistical realization from the class $G_g$ induces a unique, up to equivalence, stochastic realization from the corresponding equivalence class $S_l$ and vice versa.*

**Theorem 3.** *There is a one-to-one correspondence between the space $[S] = \{[S_l], l \in \Delta\}$ of equivalence classes $[S_l]$ of stochastic realizations and the space $[G] = \{[G_g], g \in \Gamma\}$ of equivalence* *classes of statistical realizations of the instrument. The element of* $[S]$ *corresponds to the element of* $[G]$ *if and only if they have the same sets of invariants, given by (51) and (32d), respectively. Every statistical realization from the class* $[G_g]$ *induces a unique, up to equivalence, stochastic realization from the corresponding equivalence class* $[S_l]$ *and vice versa.*

*3.4. Quantum stochastic representation of an instrument*

For further applications in the quantum measurement theory (cf. Section 4) we introduce the following notion.

**Definition.** *Let* $\mathbf{l} = \{\mathbf{b}_l, \Lambda_l, V_l\}$ *be a stochastic realization of an instrument. We shall call* $\mathbf{l}$ *quantum stochastic if for every* $\mathbf{n}$-*measurable operator-valued function* $\hat{W}_{in}^{(k)}(\mathbf{w})$ *in* (45c) *there exists a* $\mathbf{n}$-*measurable operator-valued function* $\hat{\Pi}_{in}^{(k)}(\mathbf{w})$ *such that,* $\mathbf{n}$-*a.e. on* $\Omega$, $\hat{W}_{in}^{(k)}(\mathbf{w})$ *can be represented in the factorized form:*

(52a) $$\hat{W}_{in}^{(k)}(\mathbf{w}) = \hat{\Pi}_{in}^{(k)}(\mathbf{w})\tilde{q}_{in}^{(k)}(\mathbf{w}).$$

The integral representation of an instrument through the elements of a quantum stochastic realization is given by

(52b) $$\hat{T}(E)[\hat{A}] = \sum_{i,k} \mathbf{b}_l^{(i)} \int_{\mathbf{w} \in E} \sum_{n=1}^{N(\mathbf{w})} (\hat{\Pi}_{in}^{(k)}(\mathbf{w}))^+ \hat{A} \hat{\Pi}_{in}^{(k)}(\mathbf{w}) |\tilde{q}_{in}^{(k)}(\mathbf{w})|^2 \, \tilde{\mathbf{n}}(d\mathbf{w}).$$

The following statement follows from theorem 1.

**Proposition 2.** *For any instrument there exists a quantum stochastic realization.*

**Proof.** Take some complex scalar measures $\mathbf{m}_{in}(\cdot)$, $i = 1,...,N_0$, $n = 1,...,N(\mathbf{w})$ equivalent to the positive scalar measure $\mathbf{n}(\cdot)$ in theorem 1. Let $\tilde{q}_{in}(\mathbf{w})$ be the Radon-Nykodim derivative of the measure $\mathbf{m}_{in}(\cdot)$ with respect to the measure $\mathbf{n}(\cdot)$. The $\mathbf{n}$-measurable functions $\tilde{q}_{in}(\mathbf{w})$ satisfy the relation $\tilde{q}_{in}(\mathbf{w}) \neq 0$ $\mathbf{n}$-a.e.

We can always choose the measures $\mathbf{m}_{in}(\cdot)$ in such a way that

(52c) $$\int_\Omega \sum_{n=1,...,N(\mathbf{w})} \tilde{q}_{jn}^*(\mathbf{w})\tilde{q}_{in}(\mathbf{w})\mathbf{n}(d\mathbf{w}) = \mathbf{d}_{ji}.$$

For example, we can take $\tilde{q}_{in}(\mathbf{w}) = f_i(\mathbf{w})g_n(\mathbf{w})$ with scalar complex-valued functions $f_i \in L_2(\Omega, \mathbf{n})$, $i = 1,...,N_0$ and scalar complex-valued functions $g_n \in S(\Omega, \mathbf{n})$, $n = 1,...,l$, where $l$ is equal to $\mathbf{n}$-$\sup\{N(\mathbf{w}), \mathbf{w} \in \Omega\}$, such that:
- the elements of both sets of functions $f_i(\mathbf{w}) \neq 0$, $g_n(\mathbf{w}) \neq 0$ $\mathbf{n}$-a.e.;
- the complex-valued functions $f_i(\mathbf{w})$ are mutually orthogonal $<f_i, f_j>_{L_2} = \mathbf{d}_{ij}$;
- the complex-valued functions $g_n(\mathbf{w})$ are normalized $\mathbf{n}$-a.e. on $\Omega$, that is,

(52d) $$\sum_{n=1,...,N(\mathbf{w})} |g_n(\mathbf{w})|^2 = 1.$$

Then it is easy to show that we get (45a). Introducing in (44) the operators

(52e) $$\hat{\Pi}_{in}(\mathbf{w}) = \frac{\hat{V}_{in}(\mathbf{w})}{\tilde{q}_{in}(\mathbf{w})},$$

defined $\mathbf{n}$-a.e. and satisfying the orthogonality relation

(52f) $$\int_\Omega \sum_{n=1,...,N(\boldsymbol{w})} \hat{\Pi}_{jn}^+(\boldsymbol{w})\hat{\Pi}_{in}(\boldsymbol{w})\tilde{q}_{jn}^*(\boldsymbol{w})\tilde{q}_{in}(\boldsymbol{w})\tilde{\boldsymbol{n}}(d\boldsymbol{w}) = \boldsymbol{d}_{ji}\hat{I},$$

we derive the statement of proposition 2.

If in the class $\tilde{S}_I$ there exists a quantum stochastic realization for which

(53a) $$\hat{W}_{in}^{(k)}(\boldsymbol{w}) = \hat{\Pi}^{(i)}(\boldsymbol{w})\tilde{q}_{in}^{(k)}(\boldsymbol{w}),$$

then any unitary transformation (47) of the elements of a quantum stochastic realization (53a) gives a quantum stochastic realization with different complex densities $\tilde{q}'^{(k)}_{in}(\boldsymbol{w})$ but the same (up to phase equivalence) operators $\hat{\Pi}^{(i)}(\boldsymbol{w})$. Consequently, the considered class $\tilde{S}_I$ (and $[\tilde{S}_I]$) consists of only quantum stochastic realizations of the type (53a). The operator-valued measures (48c), invariant for this class $[\tilde{S}_I]$, admit the integral representation

(53b) $$\hat{\tilde{\Theta}}_I^{(i)}(E) = \int_{\boldsymbol{w}\in E} \hat{\Pi}^{(i)}(\boldsymbol{w})\tilde{\boldsymbol{n}}_I^{(i)}(d\boldsymbol{w}), \quad \forall E \in F_B, \quad \forall i = 1,...,\tilde{N}_g,$$

which is the same for all quantum stochastic realizations of the class $[\tilde{S}_I]$.

Since for any $i = 1,...,\tilde{N}_I$ all operator-valued measures $\hat{\tilde{\Theta}}_I^{(i)}(\cdot)$ and all probability scalar measures $\tilde{\boldsymbol{n}}_I^{(i)}(\cdot)$ are invariants of the class $[\tilde{S}_I]$, the operator-valued functions $\hat{\Pi}^{(i)}(\boldsymbol{w})$, being the Radon-Nykodim derivatives $\dfrac{d\hat{\tilde{\Theta}}_I^{(i)}}{d\tilde{\boldsymbol{n}}_I^{(i)}}$, are also invariants of the class $[\tilde{S}_I]$ (up to phase equivalence). That is why, we shall use for these operators the notation $\hat{\Pi}_I^{(i)}(\boldsymbol{w})$ and call the quantum stochastic realization (53a) and the corresponding classes $\tilde{S}_I$ and $[\tilde{S}_I]$ invariant.

For the invariant class $\tilde{S}_I$ the integral representation of the instrument, corresponding to any invariant quantum stochastic realization from this class is the same and is given only through the invariants of $\tilde{S}_I$

(53c) $$\hat{T}(E)[\hat{A}] = \sum_i \boldsymbol{b}_I^{(i)} k_I(\boldsymbol{b}_I^{(i)}) \int_{\boldsymbol{w}\in E} (\hat{\Pi}_I^{(i)}(\boldsymbol{w}))^+ \hat{A} \hat{\Pi}_I^{(i)}(\boldsymbol{w})\tilde{\boldsymbol{n}}_I^{(i)}(d\boldsymbol{w})$$

Furthermore, for any invariant quantum stochastic realization from the invariant class $\tilde{S}_I$ the orthonormality relations in (45b) and (45c) for complex measures and for the operator-valued functions, respectively, can be rewritten as:

(53d) $$\int_\Omega \tilde{\boldsymbol{p}}_{ji}^{(kp)}(\boldsymbol{w})\tilde{\boldsymbol{n}}(d\boldsymbol{w}) = \boldsymbol{d}_{kp}\boldsymbol{d}_{ji}\hat{I},$$

(53e) $$\int_\Omega (\hat{\Pi}_I^{(j)}(\boldsymbol{w}))^+ \hat{\Pi}_I^{(i)}(\boldsymbol{w})\tilde{\boldsymbol{p}}_{ji}^{(kp)}(\boldsymbol{w})\tilde{\boldsymbol{n}}(d\boldsymbol{w}) = \boldsymbol{d}_{kp}\boldsymbol{d}_{ji}\hat{I},$$

with the following notation for

(53f) $$\tilde{\boldsymbol{p}}_{ji}^{(kp)}(\boldsymbol{w}) = \sum_{n=1,...,\tilde{N}(\boldsymbol{w})} (\tilde{q}_{jn}^{(k)}(\boldsymbol{w}))^* \tilde{q}_{in}^{(p)}(\boldsymbol{w}).$$

The probability scalar measures $\tilde{\boldsymbol{n}}_I^{(i)}(\cdot)$, presented by (48a) and invariant for the class $\tilde{S}_I$, have the probability densities

(53g) $$\tilde{\boldsymbol{p}}_i(\boldsymbol{w}) = \tilde{\boldsymbol{p}}_{ii}(\boldsymbol{w}),$$

(53h) $$\tilde{\boldsymbol{p}}_{ji}(\boldsymbol{w}) = (k_I(\boldsymbol{b}_I^{(i)}))^{-1} \sum_{k=1,...,k_I(\boldsymbol{b}_I^{(j)})} \tilde{\boldsymbol{p}}_{ji}^{(kk)}$$

with respect to the finite positive scalar measure $\tilde{\boldsymbol{n}}(\cdot)$.

Let $\widetilde{G}_g$ and $[\widetilde{G}_g]$ be the classes of statistical realizations, corresponding, due to theorems 2, 3 to the invariant classes $\widetilde{S}_I$ and $[\widetilde{S}_I]$, respectively. We shall call the classes of statistical realizations $\widetilde{G}_g$ and $[\widetilde{G}_g]$ invariant.

Using the notations of (43a,f) we, further, denote similar to (53f-h):

(54a) $$\boldsymbol{p}_{ji}^{(kp)}(\boldsymbol{w}) = \sum_{n=1,..,N(\boldsymbol{w})} (q_{jn}^{(k)}(\boldsymbol{w}))^* q_{in}^{(p)}(\boldsymbol{w})$$

and

(54b) $$\boldsymbol{p}_{ji}(\boldsymbol{w}) = (k_g(\boldsymbol{a}_g^{(i)}))^{-1} \sum_{k=1,...,k_g(\boldsymbol{a}_g^{(j)})} \boldsymbol{p}_{ji}^{(kk)}$$

Then, obviously, the probability densities $\boldsymbol{p}_i(\boldsymbol{w})$ of the probability scalar measure $\boldsymbol{n}_g^{(i)}(\cdot)$, $i = 1,..,N_g$, given by (43b), are

(54c) $$\boldsymbol{p}_i(\boldsymbol{w}) = \boldsymbol{p}_{ii}(\boldsymbol{w}) \geq 0$$

and the following orthonormality relations are valid:

(54d) $$\int_\Omega \boldsymbol{p}_{ji}(\boldsymbol{w})\boldsymbol{n}(d\boldsymbol{w}) = \boldsymbol{d}_{ji}, \quad \forall j,i,$$

(54e) $$\int_\Omega (\hat{\Pi}_g^{(j)}(\boldsymbol{w}))^+ \hat{\Pi}_g^{(i)}(\boldsymbol{w})\boldsymbol{p}_{ji}(\boldsymbol{w})\boldsymbol{n}(d\boldsymbol{w}) = \boldsymbol{d}_{ji}\hat{I}.$$

The following statement is valid due to theorem 2 and due to the definition of an invariant class $[\widetilde{G}_g]$.

**Theorem 4.** *Let $[\widetilde{G}_g]$ be an invariant equivalence class of statistical realizations. Then there exist*:

• *the unique family $\widetilde{\Lambda}_g$ of complex scalar measures absolutely continuous with respect to a finite positive scalar measure $\boldsymbol{n}(\cdot)$ and satisfying the orthonormality relation*:

(55a) $$\widetilde{\Lambda}_g = \{\boldsymbol{p}_{ji}(\boldsymbol{w})\boldsymbol{n}(d\boldsymbol{w}) | \boldsymbol{w} \in \Omega;\ i,j = 1,..,N_I;\ \int_\Omega \boldsymbol{p}_{ji}(\boldsymbol{w})\boldsymbol{n}(d\boldsymbol{w}) = \boldsymbol{d}_{ji}\}$$

*(the positive integer $N_g$ may be infinite)*;

• *the unique (up to phase equivalence) family $\widetilde{V}_{\tilde{a}}$ of $\boldsymbol{n}$-measurable operator-valued functions $\hat{\Pi}_g^{(i)}(\cdot)$ on $\Omega$, satisfying the orthonormality relation*,

(55b) $$\widetilde{V}_g = \{\hat{\Pi}_g^{(i)}(\boldsymbol{w}) | \boldsymbol{w} \in \Omega,\ i = 1,...,N_I,\ \int_\Omega (\hat{\Pi}_g^{(j)}(\boldsymbol{w}))^+ \hat{\Pi}_g^{(i)}(\boldsymbol{w})\boldsymbol{p}_{ji}(\boldsymbol{w})\boldsymbol{n}(d\boldsymbol{w}) = \boldsymbol{d}_{ji}\hat{I}\},$$

*and such that for any $E \in F_B$*

(55c) $$\int_{\boldsymbol{w} \in E} \hat{\Pi}_g^{(i)}(\boldsymbol{w})\boldsymbol{p}_{ii}(\boldsymbol{w})\boldsymbol{n}(d\boldsymbol{w}) \qquad i = 1,...,N_g$$

*is a bounded linear operator on $H_S$ and for any index $i$ the relation*

(55d) $$(\hat{W}_i \boldsymbol{y})(\boldsymbol{w}) = \hat{\Pi}_g^{(i)}(\boldsymbol{w})\boldsymbol{y}, \qquad \forall \boldsymbol{y} \in H_S,$$

*holding $\boldsymbol{n}$-a.e. on $\Omega$, defines the bounded linear operator $\hat{W}_i : H_S \to L_2(\Omega, \boldsymbol{n}_g^{(i)}(\cdot); H_S)$ with the norm $\|\hat{W}_i\| = 1$;*

*such that for all statistical realizations from the invariant equivalence class $[\widetilde{G}_g]$ the integral representation (41b) for any "i" instrument has the same form*:

(55e) $$\hat{T}_g^{(i)}(E)[\hat{A}] = \int_{\boldsymbol{w} \in E} (\hat{\Pi}_g^{(i)}(\boldsymbol{w}))^+ \hat{A} \hat{\Pi}_g^{(i)}(\boldsymbol{w})\boldsymbol{n}_g^{(i)}(d\boldsymbol{w})$$

*and is given only through the invariants of $[\widetilde{G}_g]$. In (55e) the probability scalar measures*

(55f) $$\boldsymbol{n}_g^{(i)}(d\boldsymbol{w}) = \boldsymbol{p}_{ii}(\boldsymbol{w})\boldsymbol{n}(d\boldsymbol{w}), \quad i = 1,...,N_g$$

*are invariants of the class* $[\widetilde{G}_g]$.

*The families* (55a) *and* (55b) *are functional invariants of the invariant class* $[\widetilde{G}_g]$.

**Definition.** *We shall call the integral representation of the instrument*

$$(56) \quad \hat{T}(E)[\hat{A}] = \sum_i \boldsymbol{a}_g^{(i)} k_g(\boldsymbol{a}_g^{(i)}) \int_{\boldsymbol{w} \in E} (\hat{\Pi}_g^{(i)}(\boldsymbol{w}))^+ \hat{A}\, \hat{\Pi}_g^{(i)}(\boldsymbol{w}) \boldsymbol{n}_g^{(i)}(d\boldsymbol{w}),$$

*corresponding to the invariant class* $\widetilde{G}_g$, a *quantum stochastic representation of the instrument*.

We would like to point out that, in general, invariants of the invariant class $\widetilde{G}_g$ such as the values of the multiplicity function $N(\boldsymbol{w}), \boldsymbol{w} \in \Omega$ and the multiplicities $k_1(\boldsymbol{a}_1^{(i)}), i = 1,\ldots, N_g$ may be greater than one.

Let now consider the description of a quantum measurement, described by the instrument represented by the quantum stochastic representation (56).

From (56) and (48a,b) it follows that the probability scalar measure (5), defining a probability distribution of outcomes, is given by

$$(57a) \quad \boldsymbol{m}_{\hat{r}_S}(E) = \sum_{i=1}^{N_g} \boldsymbol{a}_g^{(i)} k_g(\boldsymbol{a}_g^{(i)}) \int_{\boldsymbol{w} \in E} \mathrm{tr}[\boldsymbol{t}_g^{(i)}(\boldsymbol{w}, \hat{r}_S)] \boldsymbol{n}_g^{(i)}(d\boldsymbol{w})$$

through the invariants of the considered class $\widetilde{G}_g$. In (57a) we introduced the notation

$$(57b) \quad \boldsymbol{t}_g^{(i)}(\boldsymbol{w}, \hat{r}_S) = \hat{\Pi}_g^{(i)}(\boldsymbol{w}) \hat{r}_S \hat{\Pi}_g^{(i)+}(\boldsymbol{w}), \quad \forall i = 1,\ldots, N_g$$

for the unnormalized "i" statistical operators being invariants of the class $[\widetilde{G}_g]$ for the given $\hat{r}_S$. Introducing also the probability scalar measures

$$(57c) \quad \boldsymbol{m}_g^{(i)}(d\boldsymbol{w}, \hat{r}_S) = \mathrm{tr}[\boldsymbol{t}_g^{(i)}(\boldsymbol{w}, \hat{r}_S)] \boldsymbol{n}_g^{(i)}(d\boldsymbol{w})$$

for $\forall i = 1,\ldots, N_g$, we can rewrite (57a) in the form

$$(57d) \quad \boldsymbol{m}_{\hat{r}_S}(d\boldsymbol{w}) = \sum_{i=1}^{N_g} \boldsymbol{a}_g^{(i)} k_g(\boldsymbol{a}_g^{(i)}) \boldsymbol{m}_g^{(i)}(d\boldsymbol{w}, \hat{r}_S).$$

For the given initial state $\hat{r}_S$ of a quantum system, the probability scalar measures (57c) are invariants of the class $[\widetilde{G}_g]$.

The family of posterior states $\{\hat{r}(\boldsymbol{w}, \hat{r}_S), \boldsymbol{w} \in \Omega\}$, defined by (9a), is given, $\boldsymbol{m}_{\hat{r}_S}$-a.e. on $\Omega$, by the relation

$$(57e) \quad \hat{r}(\boldsymbol{w}, \hat{r}_S) = \frac{\sum_i \boldsymbol{a}_g^{(i)} k_g(\boldsymbol{a}_g^{(i)}) \boldsymbol{p}_{ii}(\boldsymbol{w}) \boldsymbol{t}_g^{(i)}(\boldsymbol{w}, \hat{r}_S)}{\sum_j \boldsymbol{a}_g^{(j)} k_g(\boldsymbol{a}_g^{(j)}) \boldsymbol{p}_{jj}(\boldsymbol{w}) \mathrm{tr}[\boldsymbol{t}_g^{(j)}(\boldsymbol{w}, \hat{r}_S)]},$$

The posterior state (57e) can be rewritten as a sum

$$(57f) \quad \hat{r}(\boldsymbol{w}, \hat{r}_S) = \sum_i \boldsymbol{q}_g^{(i)}(\boldsymbol{w}) \boldsymbol{t}_{g,N}^{(i)}(\boldsymbol{w}, \hat{r}_S)$$

of normalized statistical operators

$$(57g) \quad \boldsymbol{t}_{g,N}^{(i)}(\boldsymbol{w}, \hat{r}_S) = \frac{\boldsymbol{t}_g^{(i)}(\boldsymbol{w}, \hat{r}_S)}{\mathrm{tr}[\boldsymbol{t}_g^{(i)}(\boldsymbol{w}, \hat{r}_S)]}, \quad \forall i = 1,\ldots, N_g,$$

which are invariants for the class $[\widetilde{G}_g]$, with statistical weights

(57h) $$\boldsymbol{q}_g^{(i)}(\boldsymbol{w}) = \frac{\boldsymbol{a}_g^{(i)} k_g(\boldsymbol{a}_g^{(i)}) \boldsymbol{p}_{ii}(\boldsymbol{w}) \mathrm{tr}[\boldsymbol{\hat{t}}_g^{(i)}(\boldsymbol{w}, \hat{\boldsymbol{r}}_S)]}{\sum_j \boldsymbol{a}_g^{(j)} k_g(\boldsymbol{a}_g^{(j)}) \boldsymbol{p}_{jj}(\boldsymbol{w}) \mathrm{tr}[\boldsymbol{\hat{t}}_g^{(j)}(\boldsymbol{w}, \hat{\boldsymbol{r}}_S)]},$$

which are also invariants of the class $[\tilde{G}_g]$.

The prior (unconditional) state of a quantum system, defined by (10) in case $E = \Omega$, can be represented as

(57i)
$$\hat{\boldsymbol{r}}(\Omega, \hat{\boldsymbol{r}}_S) = \sum_{i=1}^{N_g} \boldsymbol{a}_g^{(i)} k_g(\boldsymbol{a}_g^{(i)}) \int_\Omega \boldsymbol{\hat{t}}_g^{(i)}(\boldsymbol{w}, \hat{\boldsymbol{r}}_S) \boldsymbol{n}_g^{(i)}(d\boldsymbol{w}) =$$
$$= \sum_{i=1}^{N_g} \boldsymbol{a}_g^{(i)} k_g(\boldsymbol{a}_g^{(i)}) \int_\Omega \boldsymbol{\hat{t}}_{g,N}^{(i)}(\boldsymbol{w}, \hat{\boldsymbol{r}}_S) \boldsymbol{m}_g^{(i)}(d\boldsymbol{w}, \hat{\boldsymbol{r}}_S).$$

Thus, we see that different quantum stochastic representations of the same instrument, corresponding to different invariant classes $\tilde{G}_g$ of unitarily equivalent statistical realizations, induce different decompositions (57a) and (57f) for the probability distribution of outcomes and the family of posterior states.

### 4. Quantum stochastic measurement model

The operational approach, being very important for the formalization of the complete statistical description of any generalized quantum measurement, does not, however, in general, give the possibility to include into consideration the description of the random behaviour of the quantum system under a single measurement.

However, the description of stochastic, irreversible in time behaviour of a quantum system under a single measurement is very important, in particular, in the case of continuous in time measurement, where the evolution of the continuously observed open system differs from that described by reversible in time solutions of the Schrödinger equation.

The operational approach also does not specify a description of a generalized direct quantum measurement where we have to describe the direct interaction between a classical and a quantum systems.

We would like to emphasize again that in quantum theory any physically based problem must be formulated in unitarily equivalent terms and the description of a generalized direct quantum measurement, can not be simply reduced to the quantum theory description of a measuring process.

We can not specify definitely neither the interaction, nor the quantum state of a measuring device environment, nor describe a measuring device only in quantum theory terms. In fact, under such a scheme the description of a direct quantum measurement is simply referred to the description of a direct measurement of some observable of an environment of a measuring device. Thus, the problem still remains.

We recall that for the case of discrete outcomes the original von Neumann approach [1] describes specifically a direct quantum measurement and gives both - the complete statistical description of a measurement and the complete stochastic description of the random behaviour of the quantum system under a single measurement.

Accordingly, the aim of the present section is to introduce, using the mathematical results of section 3, a new general approach, the quantum stochastic approach (QSA), to the description of a generalized direct quantum measurement, which could incorporate both the above-mentioned features of von Neumann's approach and the features of the operational approach, in the sense that this new approach would be based only on unitary invariants of a measuring process and could give the complete statistical description of a generalized direct measurement and the complete stochastic description of the random behaviour of a quantum system under a single measurement.

We consider also the description in the frame of the QSA (proposition 4) of a special kind of indirect measurement of a quantum system, where a direct measurement on some other quantum system, entangled with the one considered is described by a projection-valued POV measure.

Let us come back to the notion of a von Neumann measuring process of the observable (1), presented by the formula (17).

Fix the pair $\boldsymbol{h},\{\boldsymbol{h}_m\}$ in (17) and consider the class $G_g$ of statistical realizations of the instrument (14c), which are unitarily and phase equivalent to the statistical realization

(58a) $$\boldsymbol{g} = \{K, |\boldsymbol{h}\rangle\langle\boldsymbol{h}|, \hat{X}_g(\cdot), \hat{U}\}$$

with a projection-valued measure $\hat{X}_g(E) = \sum_{1_m \in E} |\boldsymbol{h}_m\rangle\langle\boldsymbol{h}_m|$ of the discrete type and a unitary operator $\hat{U}$, satisfying (16). The class $G_g$ has the following invariants (see (32c)):

(58b) $$[\hat{X}_g(\cdot)]; N_{X_g} = 1; \boldsymbol{a}_g = \{1\}; N_g = 1; k_g(1) = 1;$$

(58c) $$\boldsymbol{n}_g(E) = \sum_{1_m \in E} |\langle\boldsymbol{h},\boldsymbol{h}_m\rangle|^2; \quad \hat{\Theta}_g(E) = \sum_{1_m \in E} \langle\boldsymbol{h},\boldsymbol{h}_m\rangle \hat{P}_m;$$

where for any index $m$, operators $\hat{P}_m$ are projections defined by (15).

For all possible classes $G_g$ the invariants, presented in (58b), are the same. Hence, in particular, projection-valued measures $\hat{X}_g$ in statistical realizations from different classes $G_g$ are unitarily equivalent to each other (see section 3.1) and the type $[\hat{X}_g]$ is equal to the type $[\hat{P}(\cdot)]$ of the spectral projection-valued measure (14a) of a von Neumann observable (1).

Other invariants, presented in (58c), are different. In general, some scalar products $\langle\boldsymbol{h}|\boldsymbol{h}_m\rangle$ may be equal to zero and, that is why, scalar measures $\boldsymbol{n}_g(\cdot)$, corresponding to different classes $G_g$, may be of different types.

Single out classes $\tilde{G}_g$, for which $[\hat{\tilde{X}}_g] = [\tilde{\boldsymbol{n}}_g]$, what is equivalent to the condition $\langle\tilde{\boldsymbol{h}}|\tilde{\boldsymbol{h}}_m\rangle \neq 0$, $\forall m$. Every such class $\tilde{G}_g$ is invariant (cf. section 3.4) with the type $[\tilde{\boldsymbol{n}}_g(\cdot)]$ of the, corresponding to this class, invariant scalar measure $\tilde{\boldsymbol{n}}_g(\cdot)$, being equal to the type $[\hat{P}(\cdot)]$.

Moreover, only for an invariant class $\tilde{G}_g$ the type $[\tilde{\boldsymbol{n}}_g]$ of a scalar measure is defined by the type $[\hat{P}(\cdot)]$ of the spectral projection-valued measure, corresponding to the von Neumann observable (1). We recall also (cf. section 3.4) that only for an invariant class the description of a measurement can be presented through unitary invariants of a measuring process. That is why, only for an invariant class $\tilde{G}_g$ we can rewrite the expression for the instrument (14c) in the form of the quantum stochastic representation, that is, via invariants of this class - the scalar measure of discrete type $[\hat{P}(\cdot)]$ on $(R, B(R))$ and the family of operators, depending on the observed outcome, but being defined through projections in this case.

Thus, for the instrument (14c) only invariant classes of von Neumann measuring processes may be interpreted to correspond to the concept of the direct measurement of the observable (1) in the frame of original von Neumann approach.

The approach, which we introduce in this section, may be considered as the quantum stochastic generalization of the original von Neumann approach to the description of direct measurements with discrete outcomes for the case of any measurable space of outcomes, any type of a scalar measure on a space of outcomes and any type of a quantum state reduction.

Consider, for simplicity, the case when the state of a quantum system at the instant before a measurement is pure, that is, $\hat{\boldsymbol{r}}_S = |\boldsymbol{y}_0\rangle\langle\boldsymbol{y}_0|$. In this case the unnormalised statistical operators (57b), given by

(59a) $$\boldsymbol{t}_g^{(i)}(\boldsymbol{w}, \hat{\boldsymbol{r}}_S) = \hat{\Pi}_g^{(i)}(\boldsymbol{w}) |\boldsymbol{y}_0\rangle\langle\boldsymbol{y}_0| \hat{\Pi}_g^{(i)+}(\boldsymbol{w}),$$

represent pure states. The family of posterior states (57e) can be presented in the form:

(59b) $$\hat{r}(w, \hat{r}_0) = \sum_i q_g^{(i)} | \Psi_g^{(i)}(w) >< \Psi_g^{(i)}(w) |,$$

with

(59c) $$q_g^{(i)}(w) = \frac{a_g^{(i)} k_g(a_g^{(i)}) p_{ii}(w) \| \hat{\Pi}_g^{(i)}(w) y_0) \|^2}{\sum_j a_g^{(j)} k_g(a_g^{(j)}) p_{jj}(w) \| \hat{\Pi}_g^{(j)}(w) y_0) \|^2},$$

where we introduced the notation

(59d) $$\Psi_g^{(i)}(w) = \frac{\hat{\Pi}_g^{(i)}(w) y_0}{\| \Pi_g^{(i)}(w) y_0 \|_{H_S}}$$

for a normalized posterior pure state defined (up to phase equivalence) by the operator $\hat{\Pi}_g^{(i)}(w)$.

The following orthonormalization relation is valid for unnormalized posterior pure states:

(59e) $$\int_\Omega < \hat{\Pi}_g^{(j)}(w) y_0, \hat{\Pi}_g^{(i)}(w) y_0 >_{H_S} p_{ji}(w) n(dw) = d_{ji} \| y_0 \|_{H_S}^2, \quad \forall j, i, \quad \forall y_0 \in H_S.$$

From (59b) it follows that for different quantum stochastic representations of the same instrument the corresponding families of posterior pure states

(60a) $$\{\Psi_g^{(i)}(w), w \in \Omega, \ i = 1, ..., N_g\}$$

defined up to phase equivalence, and their statistical weights (59c) in the decomposition of the posterior state $\hat{r}(w, \hat{r}_0)$ are, in general, different, although the posterior state (statistical operator) $\hat{r}(w, \hat{r}_0)$ is the same.

The posterior statistical operator (see (10)), conditioned by the outcome $w \in E$, is defined by the set $\{\Psi_g^{(i)}(w)\}$ of posterior pure states as

(60b) $$\hat{r}(E, \hat{r}_S) = \frac{\sum_{i=1}^{N_g} a_g^{(i)} k_g(a_g^{(i)}) \int_{w \in E} | \Psi_g^{(i)}(w) >< \Psi_g^{(i)}(w) | m_g^{(i)}(dw)}{\sum_{i=1}^{N_g} a_g^{(i)} k_g(a_g^{(i)}) m_g^{(i)}(E)}, \quad \forall E \in F_B$$

with a probability scalar measure (57c), represented by

(60c) $$m_g^{(i)}(dw) = \| \hat{\Pi}_g^{(i)}(w) y_0) \|^2 n_g^{(i)}(dw).$$

The prior state (57i) has the form

(60d) $$\hat{r}(\Omega, \hat{r}_S) = \sum_{i=1}^{N_g} a_g^{(i)} k_g(a_g^{(i)}) \int_\Omega | \Psi_g^{(i)}(w) >< \Psi_g^{(i)}(w) | m_g^{(i)}(dw)$$

and can be considered as the usual statistical average over the posterior pure states (60a) with respect to the probability scalar measure $m_g^{(i)}(dw)$ in the "$i$" random channel of measurement and with respect to the different channels, given with the statistical weights $a_g^{(i)} k_g(a_g^{(i)}), \cdots \forall i = 1, ..., N_g$, $\sum_i a_g^{(i)} k_g(a_g^{(i)}) = 1$.

From (60b,d) it follows that $\Psi_g^{(i)}(w)$ can be interpreted as a random posterior pure state outcome in a Hilbert space $H_S$ of a quantum system, conditioned by the observed value $w \in dw$, in the "$i$" random channel. For the definite $w$ the probabilities of different posterior pure state outcomes are defined by (59c). We can interpret then the probability scalar measure $n_g^{(i)}(dw)$ in (60c) as describing the input probability distribution of different outcomes in the "$i$" channel and the scalar measure $\mathbf{m}_g^{(i)}(d\mathbf{w})$ as describing the output probability distribution in the "$i$" channel of a given invariant quantum stochastic representation.

Thus, the random operators $\hat{\Pi}_g^{(i)}(\mathbf{w})$ can be interpreted as describing under a single quantum measurement the stochastic behaviour of a quantum system, conditioned by the observed outcome $\mathbf{w} \in d\mathbf{w}$ in the "$i$" random measurement channel.

Analysing the definition of the invariant class of unitarily equivalent statistical realizations and the description, given by (59)-(60), of the probability distribution of outcomes and the family of posterior states, corresponding to this class, we conclude that different quantum stochastic representations of the same instrument can be identified with the description of different generalised direct quantum measurements.

Although the statistical description of these measurements (the POV measure and the family of normalized posterior states) is the same, the stochastic behaviour of a quantum system in the sense of specification of the probabilistic transition law governing the change from the initial state of the quantum system to a final one under a single measurement, may be different.

Physically, the notion of different channels under a direct measurement corresponds under the same observed outcome of a measured quantum variable to different underlying random quantum transitions of the environment of a measuring device, which we can not, however, specify with certainty.

The following proposition follows from our identification of the description of an invariant class of unitarily and phase equivalent statistical realizations with a concrete direct quantum measurement.

**Proposition 3.** *For any generalized direct quantum measurement with outcomes in a standard Borel space ( $\Omega, F_B$ ) upon a quantum system being at the instant before the measurement in a state $\hat{\mathbf{r}}_S$, there exist*:

• *the unique family of complex scalar measures, absolutely continuous with respect to a finite positive scalar measure $\mathbf{n}(\cdot)$ and satisfying the orthonormality relation*:

(61) $$\Lambda = \{\mathbf{p}_{ji}(\mathbf{w})\mathbf{n}(d\mathbf{w}) \mid \mathbf{w} \in \Omega; \, i,j = 1,2,..,N_0; \int_\Omega \mathbf{p}_{ji}(\mathbf{w})\mathbf{n}(d\mathbf{w}) = \mathbf{d}_{ji}\},$$

*where the integer $N_0$ may be infinite*;

• *the unique (up to phase equivalence) family of $\mathbf{n}$- measurable operator-valued functions $\hat{V}_i(\cdot)$ on $\Omega$, satisfying the orthonormality relation with respect to scalar measures* (61):

(62a) $$V = \{\hat{V}_i(\mathbf{w}) \mid \mathbf{w} \in \Omega; \, i = 1,2,..,N_0; \int_\Omega \hat{V}_j^+(\mathbf{w})\hat{V}_i(\mathbf{w})\mathbf{p}_{ji}(\mathbf{w})\mathbf{n}(d\mathbf{w}) = \mathbf{d}_{ji}\hat{I}\}$$

*and such that for any index $i = 1,...,N_0$ and for $\forall E \in F_B$*

(62b) $$\int_{\mathbf{w} \in E} \hat{V}_i(\mathbf{w})\mathbf{n}_i(d\mathbf{w})$$

*is a bounded linear operator on $H_S$. The relation*

(62c) $$(\hat{W}_i \mathbf{y})(\mathbf{w}) = \hat{V}_i(\mathbf{w})\mathbf{y}, \qquad \forall \mathbf{y} \in H_S,$$

*holding $\mathbf{n}$-a.e. on $\Omega$, defines the bounded linear operator $\hat{W}_i : H_S \to L_2(\Omega, \mathbf{n}_i; H_S)$ with the norm $\|\hat{W}_i\| = 1$. In (62a,b,c)*

(62d) $$\mathbf{n}_i(d\mathbf{w}) = \mathbf{p}_{ii}(\mathbf{w})\mathbf{n}(d\mathbf{w});$$

• *the unique sequence of positive numbers $\mathbf{a} = (\mathbf{a}_1, \mathbf{a}_2,...,\mathbf{a}_{N_0}), N_0 \leq \infty$, satisfying the relation*

(63) $$\sum_{i=1}^{N_0} \mathbf{a}_i = 1;$$

*such that the complete statistical description (a POV measure and a family of posterior states) of a measurement and the complete stochastic description of the behaviour of a quantum system under a single measurement (a family of posterior pure state outcomes and their probability distribution) are given by:*

- *The POV measure*

(64) $$\hat{M}(E) = \sum_i \bm{a}_i \int_{\bm{w} \in E} \hat{V}_i^+(\bm{w}) \hat{V}_i(\bm{w}) \bm{n}_i(d\bm{w}), \quad \forall E \in F_B,$$

- *The family $\{\hat{\bm{r}}(\bm{w}, \hat{\bm{r}}_S), \bm{w} \in \Omega\}$ of posterior states*

(65a) $$\hat{\bm{r}}(\bm{w}, \hat{\bm{r}}_S) = \sum_i \bm{x}_i(\bm{w}) \bm{t}_i(\bm{w}, \hat{\bm{r}}_S)$$

*with*

(65b) $$\bm{t}_i(\bm{w}, \hat{\bm{r}}_S) = \hat{V}_i(\bm{w}) \hat{\bm{r}}_S \hat{V}_i^+(\bm{w}), \quad \bm{x}_i(\bm{w}) = \frac{\bm{a}_i \bm{p}_{ii}(\bm{w})}{\sum_j \bm{a}_j \bm{p}_{jj}(\bm{w}) \mathrm{tr}[\bm{t}_j(\bm{w}, \hat{\bm{r}}_S)]};$$

- *The probability scalar measure of the whole measurement, given by the expression*

(66a) $$\bm{m}_{\bm{r}_S}(\cdot) = \sum_i \bm{a}_i \bm{m}_{\bm{r}_S}^{(i)}(\cdot)$$

*through the probability scalar measures*

(66b) $$\bm{m}_{\bm{r}_S}^{(i)}(d\bm{w}) = \mathrm{tr}[\bm{t}_i(\bm{w}, \hat{\bm{r}}_S)] \bm{n}_i(d\bm{w})$$

*in different "i" random channels of a measurement;*

- *The family of random operators (62), describing the stochastic behaviour of the quantum system under a single direct measurement. Every operator $\hat{V}_i(\bm{w})$ defines (up to phase equivalence) in the Hilbert space $H_S$ a posterior pure state outcome conditioned by the observed result $\bm{w} \in d\bm{w}$ in the "i" random channel of a measurement. For any $\bm{y}_0 \in H_S$ the following orthonormality relation for a family $\{\hat{V}_i(\bm{w})\bm{y}_0, \bm{w} \in \Omega, i = 1,..., N_0\}$ of unnormalized posterior pure state outcomes is valid:*

(67a) $$\int_\Omega <\hat{V}_j(\bm{w})\bm{y}_0, \hat{V}_i(\bm{w})\bm{y}_0>_{H_S} \bm{p}_{ji}(\bm{w})\bm{n}(d\bm{w}) = \bm{d}_{ji} \|\bm{y}_0\|_{H_S}^2, \quad \forall j,i.$$

*The probability distribution of different outcomes $\bm{w}$ in a random "i" channel is presented by (66b). The statistical weights of different random channels of a measurement are given by numbers $\bm{a}_i, i = 1,..., N_0$. For the definite observed outcome $\bm{w}$ the probability of the posterior pure state outcome $\hat{V}_i(\bm{w})\bm{y}_0$ is given by*

(67b) $$\bm{q}_i(\bm{w}) = \frac{\bm{a}_i \bm{p}_{ii}(\bm{w}) \|\hat{V}_i(\bm{w})\bm{y}_0\|^2}{\sum_j \bm{a}_j \bm{p}_{jj}(\bm{w}) \|\hat{V}_j(\bm{w})\bm{y}_0\|^2}$$

*The prior state*

(67c) $$\hat{\bm{r}}(\Omega, \hat{\bm{r}}_S) = \sum_{i=1}^{N_0} \bm{a}_i \int_\Omega |\hat{V}_i(\bm{w})\bm{y}_0><\hat{V}_i(\bm{w})\bm{y}_0| \bm{n}_i(d\bm{w}) =$$
$$= \sum_{i=1}^{N_0} \bm{a}_i \int_\Omega |\Psi_g^{(i)}(\bm{w})><\Psi_g^{(i)}(\bm{w})| \bm{m}_{\bm{r}_S}^{(i)}(d\bm{w}),$$
$$\Psi_g^{(i)}(\bm{w}) = \frac{\hat{V}_i(\bm{w})\bm{y}_0}{\|\hat{V}_i(\bm{w})\bm{y}_0\|}$$

*is a statistical average over the posterior pure state outcomes with respect to the probability distribution of outcomes in every " i " random measurement channel and with respect to the different measurements channels.*

We shall call $\hat{V}_i(\mathbf{w})$ a quantum stochastic evolution operator and the probability scalar measures $\mathbf{n}_i(\cdot)$, $\mathbf{n}_0(\cdot) = \sum_i \mathbf{a}_i \mathbf{n}_i(\cdot)$ and $\mathbf{m}_{T_S}^{(i)}(\cdot)$, $\mathbf{m}_{T_S}(\cdot)$ as input and output probability measures, respectively. We shall also call the triple $\{\Lambda, V, \mathbf{a}\}$ a quantum stochastic representation of a generalized direct quantum measurement.

Direct measurements, presented by different quantum stochastic representations, are called stochastic representation equivalent if the complete statistical and complete stochastic description of these measurements is identical. In the frame of the quantum stochastic approach projective direct measurements present such a stochastic representation equivalence class of direct measurements on $(R, B(R))$, for which the complete statistical and the complete stochastic description is given by the von Neumann measurement postulates [1], presented by the formulae (2), (3).

Consider now also the case of indirect measurement. The following proposition is a corollary of theorem 2.

**Proposition 4.** Let $H_S$ be a Hilbert space of a quantum system and $(\Omega, F_B)$ be a standard Borel space $(\Omega, F_B)$. *For any collection, consisting of*:

• *a sequence of positive coefficients*

(68a) $$\mathbf{b} = (\mathbf{b}_1, ...., \mathbf{b}_{N_0}), \ N_0 \leq \infty, \ \sum_{i=1}^{N_0} \mathbf{b}_i = 1;$$

• *a family $\Lambda$ of complex scalar measures, absolutely continuous with respect to a finite positive scalar measure $\mathbf{n}(\cdot)$ and satisfying the orthonormality relation*:

(68b) $$\Lambda = \{q_{in}(\mathbf{w})\mathbf{n}(d\mathbf{w}) | \mathbf{w} \in \Omega; \ i = 1,..,N_0; \ n = 1,..,N(\mathbf{w}); \int_\Omega \sum_{n=1}^{N(\mathbf{w})} q_{jn}^*(\mathbf{w})q_{in}(\mathbf{w})\mathbf{n}(d\mathbf{w}) = \mathbf{d}_{ji}\},$$

*where the positive integer $N(\mathbf{w})$ may be infinite;*

• *a family $V$ of $\mathbf{n}$-measurable operator-valued functions $\hat{V}_{in}(\cdot)$ on $\Omega$, satisfying the orthonormality relation*:
(68c)
$$V = \{\hat{V}_{in}(\mathbf{w}) | \mathbf{w} \in \Omega, \ i = 1,...,N_0; n = 1,...N(\mathbf{w}); \int_\Omega \sum_{n=1}^{N(\mathbf{w})} \hat{V}_{jn}^+(\mathbf{w})\hat{V}_{in}(\mathbf{w})q_{jn}^*(\mathbf{w})q_{in}(\mathbf{w})\mathbf{n}(d\mathbf{w}) = \mathbf{d}_{ji}\hat{I}\};$$

*and such that for any indexes $i, n$ and any $E \in F_B$*

(68d) $$\int_{\mathbf{w} \in E} \hat{V}_{in}(\mathbf{w}) q_{in}(\mathbf{w}) \mathbf{n}(d\mathbf{w}), \quad \int_{\mathbf{w} \in E} \sum_{n=1,..., N(\mathbf{w})} \hat{V}_{in}(\mathbf{w}) | q_{in}(\mathbf{w})|^2 \mathbf{n}(d\mathbf{w})$$

*are bounded linear operators on $H_S$;*

*there exists an indirect measurement of a quantum system, induced by a direct measurement upon some other quantum system, described by a Hilbert space $H_R$ and entangled with the one considered, such that the POV measure $\hat{P}(\cdot): F_B \to L(H_R)$ of this direct measurement is projection-valued and consistent with the triple, given by (68b,c).*
*The instrument, corresponding to this indirect measurement, is given by*

(68e) $$\hat{T}(E)[\hat{A}] = \sum_i \mathbf{b}_i \int_{\mathbf{w} \in E} \sum_{n=1}^{N(\mathbf{w})} \hat{V}_{in}^+(\mathbf{w}) \hat{A} \hat{V}_{in}(\mathbf{w}) | q_{in}(\mathbf{w})|^2 \mathbf{n}(d\mathbf{w}).$$

In Proposition 4 the projection-valued measure $\hat{P}(\cdot)$ is said to be *consistent* with the families (68b,c) if the type $[\hat{P}(\cdot)]$ is equal to $[\boldsymbol{n}]$ and the multiplicity function $N_P$ of $\hat{P}(\cdot)$ is equal to $N(\boldsymbol{w})$ $\boldsymbol{n}$-a.e. on $\Omega$.

## 5. Semiclassical stochastic model of a quantum measurement

In quantum theory there was always a wish to combine the classical description of a measuring apparatus for an observer with the quantum description of an observed system.

The results we derived in the previous sections allow us to introduce such kind of interpretation of the description of a generalized direct quantum measurement.

**Definition** *(of a classical premeasurement state of a measuring device). We shall say that a family of scalar measures*

(69a) $$\Lambda_0 = \{\boldsymbol{p}_{ji}(\boldsymbol{w})\boldsymbol{n}(d\boldsymbol{w}) \mid \boldsymbol{w} \in \Omega; \, i, j = 1, 2, .., N_0; \int_\Omega \boldsymbol{p}_{ji}(\boldsymbol{w})\boldsymbol{n}(d\boldsymbol{w}) = \boldsymbol{d}_{ji}\}$$

*on a measurable space* $(\Omega, F)$ *describes a classical premeasurement state* $\Lambda_0$ *of a quantum apparatus if for any measurement* "$\boldsymbol{a}$":

(69b) $$\boldsymbol{a} = (\boldsymbol{a}_1, \boldsymbol{a}_2, ....), \, \boldsymbol{a}_i > 0, \, \sum_i \boldsymbol{a}_i = 1,$$

*performed by a "free" apparatus in a state* $\Lambda_0$*, a probability scalar measure of* "$\boldsymbol{a}$" *measurement is given by*

(69c) $$\boldsymbol{n}_0^{(\boldsymbol{a})}(d\boldsymbol{w}) = \sum_i \boldsymbol{a}_i \boldsymbol{p}_{ii}(\boldsymbol{w})\boldsymbol{n}(d\boldsymbol{w}).$$

Physically different "$\boldsymbol{a}$" correspond to different preparations of the quantum state of a measuring apparatus.

Let at the instant before a measurement a measuring device be in a classical premeasurement state $\Lambda_0$ and a quantum system be in a quantum state $\hat{\boldsymbol{r}}_S$. From this above definition and proposition 3 it follows that for any measurement "$\boldsymbol{a}$", performed by a measuring device upon a quantum system, there exists the unique (up to phase equivalence) family of quantum stochastic evolution operators (62) such that the complete statistical and the complete stochastic description of a measurement is given by formulae (64)-(67).

## 6. Concluding remarks

In the present paper we review the main approaches to the description of quantum measurements. We analyse the structure of different classes of statistical and stochastic realizations of an instrument, find their invariants and introduce the notion of a quantum stochastic representation of an instrument, whose elements are wholly determined by the invariants of the corresponding invariant class of unitarily and phase equivalent statistical realizations.

We show that the description of a generalized direct quantum measurement can be considered in the frame of a new general approach (QSA), based on the notion of a family of quantum stochastic evolution operators, satisfying the orthonormality relation and describing under a measurement the conditional evolution of a quantum system in a Hilbert space.

The proposed approach allows to give:
- the complete statistical description (a POV measure and a family of posterior states) of any generalized direct quantum measurement;
- the complete description in a Hilbert space of the stochastic behaviour of a quantum system under a generalized direct measurement in the sense of specification of the probabilistic transition law governing the change from the initial state of a quantum system to a final one under a single measurement;

- to give the semiclassical interpretation of the description of a quantum measurement;
- to formalize the consideration of all possible cases of generalized direct quantum measurements, including measurements continuous in time.

In a sequel to this paper we shall consider in detail the further application of the proposed general approach to the description of different concrete types of measurements.

**Acknowledgements**

I am very grateful for the hospitality of the Centre for Mathematical Physics and Stochastics, University of Aarhus, Denmark. I am greatly indebted to Professor Ole E. Barndorff-Nielsen for helpful discussions, for his valuable remarks and for support of the work on this paper. Very detailed and helpful comments from the viewers are much appreciated.